\documentclass[twocolumn,showpacs,preprintnumbers]{revtex4-1}

\usepackage[english]{babel} 
\usepackage{graphics} 
\usepackage{graphicx} 
\usepackage{amssymb} 
\usepackage{amsthm} 
\usepackage{amsmath} 
\usepackage{siunitx} 
\usepackage{subfig} 

\begin{document}


\title{Global phenomenological descriptions of nuclear odd-even mass staggering}
\author{D. Hove, A.S. Jensen, K. Riisager}
\affiliation{Department of Physics and Astronomy, Aarhus University, DK-8000 Aarhus C, Denmark} 
\date{\today}


\begin{abstract}
We examine the general nature of nuclear odd-even mass differences by
employing neutron and proton mass relations that emphasize these
effects. The most recent mass tables are used.  The possibility of a
neutron excess dependence of the staggering is examined in detail in
separate regions defined by the main nuclear shells, and a clear
change in this dependency is found at $Z=50$ for both neutrons and
protons.  A further separation into odd and even neutron (proton)
number produces very accurate local descriptions of the mass
differences for each type of nucleons.  These odd-even effects are
combined into a global phenomenological expression, ready to use in a
binding energy formula.  The results deviate from previous
parametrizations, and in particular found to be significantly superior
to a recent two term, $A^{-1}$ dependence.
\end{abstract}


\maketitle



\section{Introduction \label{Introduction}} 

The systematic odd-even mass difference in nuclei was recognized early
on, see \cite{boh75} for a brief historical overview. Several effects
will contribute to the experimentally observed differences, in
particular pairing and the twofold degeneracy of orbits, see
\cite{Dea03,ber09,ber09a} for an overview of recent work.
Traditionally the odd-even mass staggering has been parametrized by a
power law in the mass number $A$ (both $A^{-1/2}$ and $A^{-1/3}$ have
been used), but other functional forms have been used, e.g.\ a
constant and a $A^{-1}$ term in \cite{ber09}.  We emphasize that these
smoothly varying parametrizations only were meant to reproduce values
of staggering averaged over many neighboring nuclei.  The odd-even
mass differences has traditionally been attributed to nucleon pairing
(the largest part) and breaking of the time reversal double degeneracy
of the single particle levels (the smallest part).  Other effects may
contribute as well.

Both experimentally and theoretically one observes significant,
systematic deviations from the simple laws \cite{ber09a} and it
therefore seems worthwhile to make use of the recently updated,
extensive and accurate, nuclear mass table \cite{Audi} to look more
carefully for trends in the experimental odd-even staggering.  We
shall in particular reinvestigate the suggestion of an explicit
dependence on neutron excess \cite{Vog84,Aksel} that was not supported
by nuclear pairing models \cite{Mol92}. Our aim is to find an improved
phenomenological description of the odd-even mass differences that may
be used in combination with semi-empirical mass models, and perhaps
reveal trends that could inspire future more basic theoretical work.

The relevant theoretical considerations are presented in
Sec.~\ref{Sect. theory}. In particular, the relevant mass relations
designed specifically to isolate odd-even effects are introduced. Mass
relations of this nature were investigated in detail by Jensen
\textit{et al.} \cite{Aksel}, but the relations used here are more
compact, and are applied with the sole purpose of analyzing odd-even
effects in general.

Section~\ref{Sect. Sys} contains the initial examination of odd-even
mass differences.  The focus is on the general structure of the
staggering effects, and to that end the results, free of any
manipulations, are presented in this section. To provide a more
general overview of this structure Sec.~\ref{Sec. E-O} presents a
three dimensional illustration of neutron staggering as a function of
both $N$ and $Z$. Also included in Sec.~\ref{Sect. Sys} is a short
evaluation of the extent of the shell effects.

Section \ref{Sec. Gen} contains the examination of staggering effects
as a function of isospin projection.  This includes both a separation
according to odd-even neutron and proton configurations, as well as
separation into regions defined by nuclear shells. Having established
the effect of each separation the results are combined into one global
expression, which describes the collective odd-even staggering effect
and includes neutron-proton pairing explicitly. Finally, in the results are compared to a very
accurate recent two-term description with an $A^{-1}$ dependency.



\section{Theoretical foundation \label{Sect. theory}}

Fundamental to all following examinations is a rather general
separation of the binding energy into three parts \cite {bra72}
\begin{align}
B(N,Z) = B_{LD}(N,Z) + B_{sh}(N,Z) + \Delta(N,Z).  \label{B}
\end{align}
All smooth aspects are contained in the liquid drop term, $B_{LD}$,
whereas $B_{sh}$ accounts for the smaller, but faster oscillating
localized, nuclear shells.  The last term, $\Delta$, has to include
all other contributions to the nuclear binding energy, that is various
types of correlations and in particular variations depending on the
parity of the nucleon numbers.  The intent in the present paper is to
analyze the neutron and proton staggering included in $\Delta$ looking
for possible global dependencies.  To that end any influence from both
$B_{LD}$ and $B_{sh}$ must be eliminated.  Since these terms by far
are the largest some care must be exercised to isolate the desired
effects from the measured binding energies.

The smooth contributions (mainly the liquid drop term, $B_{LD}$) is in
principle easily eliminated to any order by appropriately conctructed
binding energy differences \cite{Aksel}.  Removal of first order is
achieved by use of the double difference
\begin{align}
Q(n, z) 
=& -B(N-n, Z-z) + 2 B(N,Z) \notag \\
&- B(N+n, Z+z). \label{Q}
\end{align}
Taylor expansion of the smooth part, $B_{LD}$, of each of the three
$B-$terms define $Q_{LD}$ as the smooth part of $Q$, that is
\begin{align}
Q_{LD} = -  n^2 \frac{\partial^2 B_{LD}}{\partial N^2}
- 2 nz \frac{\partial^2 B_{LD}}{\partial N \partial Z} - z^2 \frac{\partial^2 B_{LD}}{\partial Z^2}.~\label{Q_LD}
\end{align}
This reduction, to only second order contributions of smooth part
through Eq.~(\ref{Q}), is very significant but it might not provide
sufficient accuracy.  Extension to more elaborate mass relations would
formally improve this accuracy, but introduce other uncertainties.  We
shall instead use the liquid drop model itself to eliminate remaining
smooth contributions.  This amounts to use of $Q-Q_{LD}$ instead of
$Q$, where $Q_{LD}$ can be taken from Eq.~(\ref{Q_LD}) or directly
from a liquid drop model without any expansion. Our method to extract the odd-even staggering eliminates almost all liquid drop smooth background contributions. As our intention is to study the odd-even staggering specifically, and not the liquid drop model itself, a simple version is employed containing only the most fundamental terms
\begin{align}
B_{LD} = -a_v A + a_s A^{2/3} + a_c \frac{Z^2}{A^{1/3}} + a_a \frac{(Z- A/2)^2}{A},
\end{align}
where $a_v = 15.56$, $a_s = 17,23$, $a_c = 0.697$, and $a_a = 93.14$,
all in units of $\si{\mega\electronvolt}$. The accuracy with this expression and those parameter values is sufficient for our purpose. 

Any mass relation based on Eq.~(\ref{Q}), and with the liquid drop
terms subtracted, will then dramatically reduce contributions from the
unwanted smooth parts of the binding energy.  We also need to
eliminate unwanted contributions from $B_{sh}$.  Shell effects vary
rather discontinuously across magic numbers while relatively smooth by
moving small steps from magic number to either side of it.  Therefore
the three-point mass relation in Eq.~(\ref{Q}) would have only a very
small contribution from $B_{sh}$ provided magic numbers for both $N$
and $Z$ are excluded.  In general only one side of magic numbers
should be allowed to enter the mass relations employed.  These
expectations will be examined in greater detail in
Sec.~\ref{Sec. Shell}.

\begin{figure}
\centering

\includegraphics[width=1.0\columnwidth]{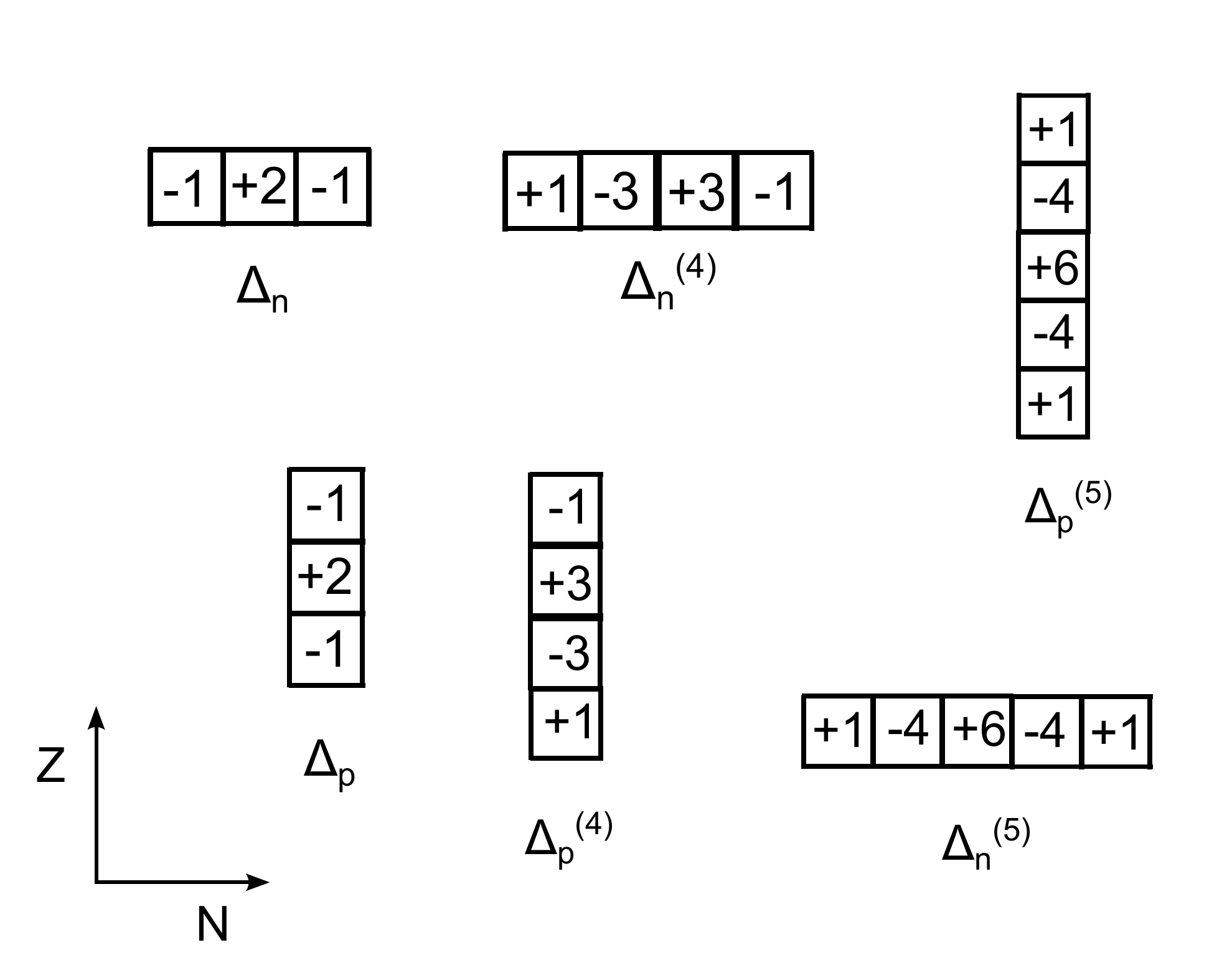}

\caption{The structure of $\Delta_{n}$, $\Delta_p$, $\Delta_{n}^{(4)}$, $\Delta_{p}^{(4)}$, $\Delta_{n}^{(5)}$, and $\Delta_{p}^{(5)}$ given in the NZ-plane. The relative weight of each individual isotope, not including normalizing factors, is also included. \label{Struc.}}
\end{figure}

Choosing a specific mass relation which emphasizes either neutron or
proton odd-even staggering now allows detailed and accurate
investigations of this effect contained in $\Delta$ from
Eq.~(\ref{B}).  Two mass relations are in particular ideally suited to
study the neutron and proton staggering. They are given as
\begin{align}
\Delta_n = \frac{1}{2} \pi_n \left( Q(1,0) - Q_{LD}(1,0) \right), \label{Dn}\\
\Delta_p = \frac{1}{2} \pi_p\left(Q(0,1) - Q_{LD}(0,1) \right), \label{Dp}
\end{align}
where $\pi_n$ and $\pi_p$ assure a positive result, when defined as
\begin{align}
\pi_n &= (-1)^{N}, &\pi_p = (-1)^{Z}.
\end{align}
The definition and normalization of $\Delta_n$ corresponds to an
additional binding energy of $\Delta_n$ for even compared to odd
values of $N$.  The nuclei contributing to $\Delta_n$ and $\Delta_p$
in the NZ-plane are seen in Fig.~\ref{Struc.}.  They are ideal for
isolating odd-even effects as either horizontal or vertical with
alternating sign for each isotope.  When we have extracted $\Delta_n$
and $\Delta_p$ from experimental masses, either as numbers or as
parametrized analytic expressions, the corresponding
contribution to $\Delta$ in Eq.~(\ref{B}) could be expected to be
\begin{align}
\Delta = \frac{1}{2} \left(\pi_n\Delta_n +\pi_p \Delta_p \right). \label{Delta}
\end{align}
This expression has the classical form for the odd-even mass
difference, often referred to as pairing effects although other
substantial contributions also can be included. 

It may appear as if $\Delta_{n}$ and $\Delta_{p}$ are expressions
related solely to either neutron or proton odd-even mass differences.
However, the three-point mass relation in Eq.~(\ref{Q}) also includes
a contribution from neutron-proton pairing effects that are known to
be sizable \cite{Aksel,Friedman}.  Assuming the neutron-proton pairing
effect results in a term, $C\pi_{np} = C (1-\pi_n)(1-\pi_p)/4$ (i.e.\
only a contribution for odd-odd nuclei), then $\Delta_n$ would receive
a contribution $-C(1-\pi_p)/2$, and analogously for $\Delta_p$ and
Eq.~(\ref{Delta}) must be corrected for this.  Equations (\ref{Dn})
and (\ref{Dp}) therefore reflect odd-even effects in general,
including terms arising from possible neutron-proton couplings.

It is a choice to use three-point mass relations to study odd-even
effects.  In fact, any number of neighboring masses can be combined to
provide information about similar effects. Still in any case, unwanted
contributions must be eliminated.  Replacing the binding energies in
Eq.~(\ref{Q}) with separation energies of the form $S(N,Z) = B(N,Z) -
B(N-n^{\prime},Z-z^{\prime})$ would eliminate the smooth terms to
second order, as demonstrated by Jensen \textit{et al.} \cite{Aksel}.
The corresponding mass relations with $n=z=n'=z'=1$ result in a
combination of four nuclei, denoted $\Delta_n^{(4)}$ and
$\Delta_p^{(4)}$ as shown in Fig.~\ref{Struc.}.  This type of
four-point nuclear mass relation contains unequal weights on even and
odd nuclei which only results in minor inaccuracies.  It is ideal for
studying the neutron and proton pairing, but it will average the
results for odd and even particle numbers and is therefore not
recommended \cite{ber09,Sat98}.

Alternatively, a structure eliminating smooth terms to third order
could be constructed. This would combine five nuclei as demonstrated
by $\Delta_{n,p}^{(5)}$ in Fig.~\ref{Struc.}. The main drawback is the
extent of the structure which implies averaging over more nuclei
further apart.  By combining five nuclei relatively far apart the
odd-even effects would be significantly diminished as a result of the
implicit averaging, see also the detailed comparison between $\Delta$
and $\Delta^{(5)}$ in ref. \cite{Dug02}.

In summary, for investigation of odd-even mass staggering the most
suitable structure is the compact, three nuclei structure with the
liquid drop contribution subtracted as presented in Eqs.~(\ref{Dn})
and (\ref{Dp}).
 
In general, the term $\Delta$ in $B$ from Eq.~(\ref{B}) by definition
contains all the effects beyond those included in $B_{LD}$ and
$B_{sh}$.  Extraction of specific contributions to the nuclear binding
energy is done by construction of mass relations dedicated to isolate
the desired effects and simultaneously remove all significant
contributions from $B_{LD}$ and $B_{sh}$.  These two requirements are
not altogether mutually compatible. Removal of smooth parts in $B$
cannot distinguish between the different terms, $B_{LD}$,
$B_{sh}$ or $\Delta$.  Only the form of the parts of the assumed
$\Delta$ contribution is distinguishable, but within such a form still
smooth contributions would vanish, even when it is desirable to know
them.

We emphasize that extraction of each type of correlation contribution
has to be done with a precisely corresponding mass relation. The
result is an additive piece (not everything) to the $\Delta$ term in
$B$ which then should be included in future binding energy
expressions.  A number of different terms can then accumulate.

\section{Examining odd-even neutron and proton staggering \label{Sect. Sys}} 

In the following section the mass relations defined in Eq.~(\ref{Dn})
and (\ref{Dp}) are used to examine neutron and proton mass staggering
respectively. In Sec.~\ref{Sec. E-O} a detailed look at the neutron
effect in three dimensions is presented. This should provide a more
general insight into the nature and structure of the odd-even
effects. Finally, in Sec.~\ref{Sec. Shell} the extent of the shell
effects is examined based on work by Dieperink and Van Isacker
\cite{Diep}.  All measurements used are taken from the recent
compilation by G. Audi \textit{et al.} \cite{Audi}.

\subsection{Isolated odd-even effects}

\begin{figure}
\centering

\subfloat
{
   \includegraphics[width=1\columnwidth]{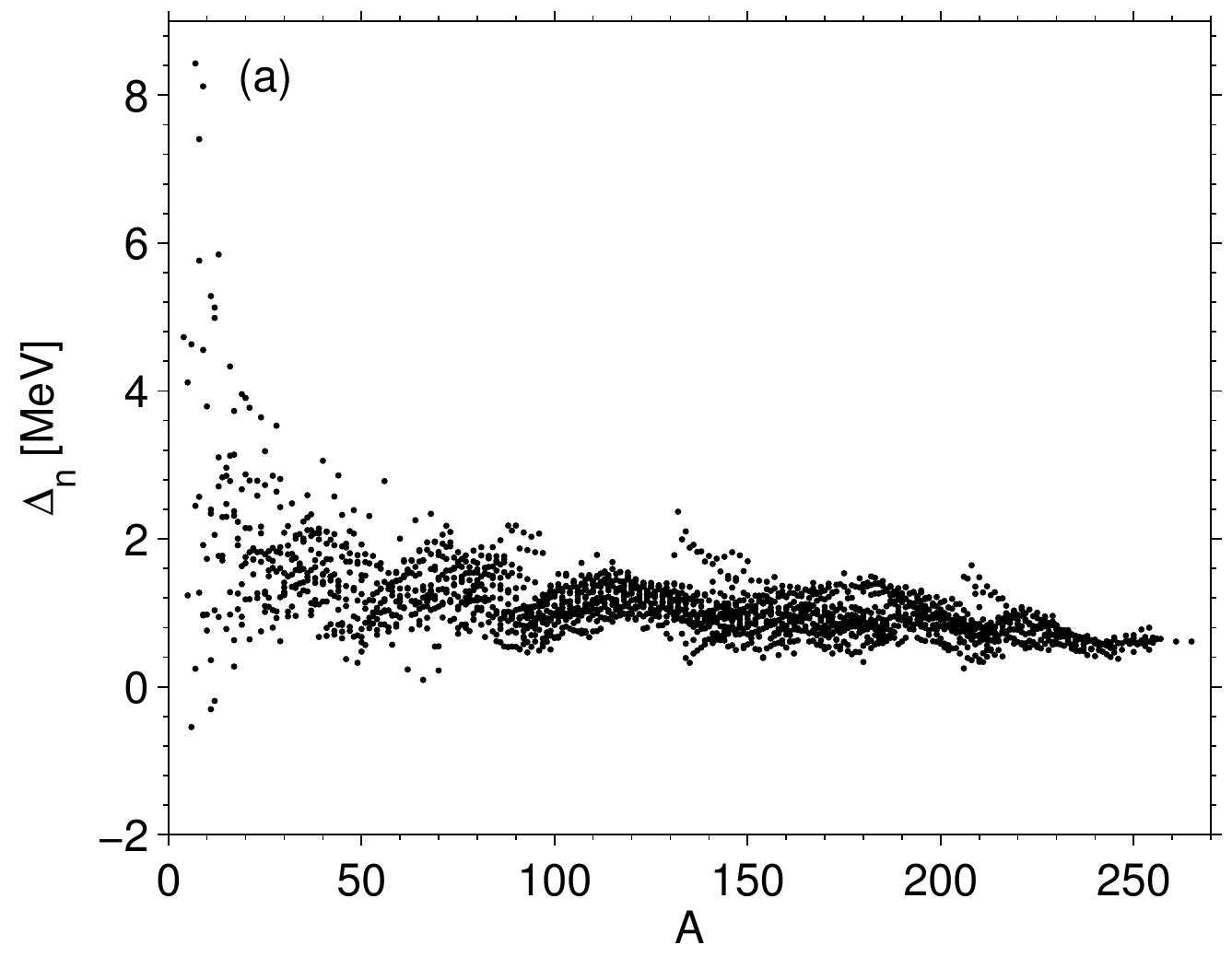} \label{pure n}
 }

\subfloat
{
   \includegraphics[width=1\columnwidth]{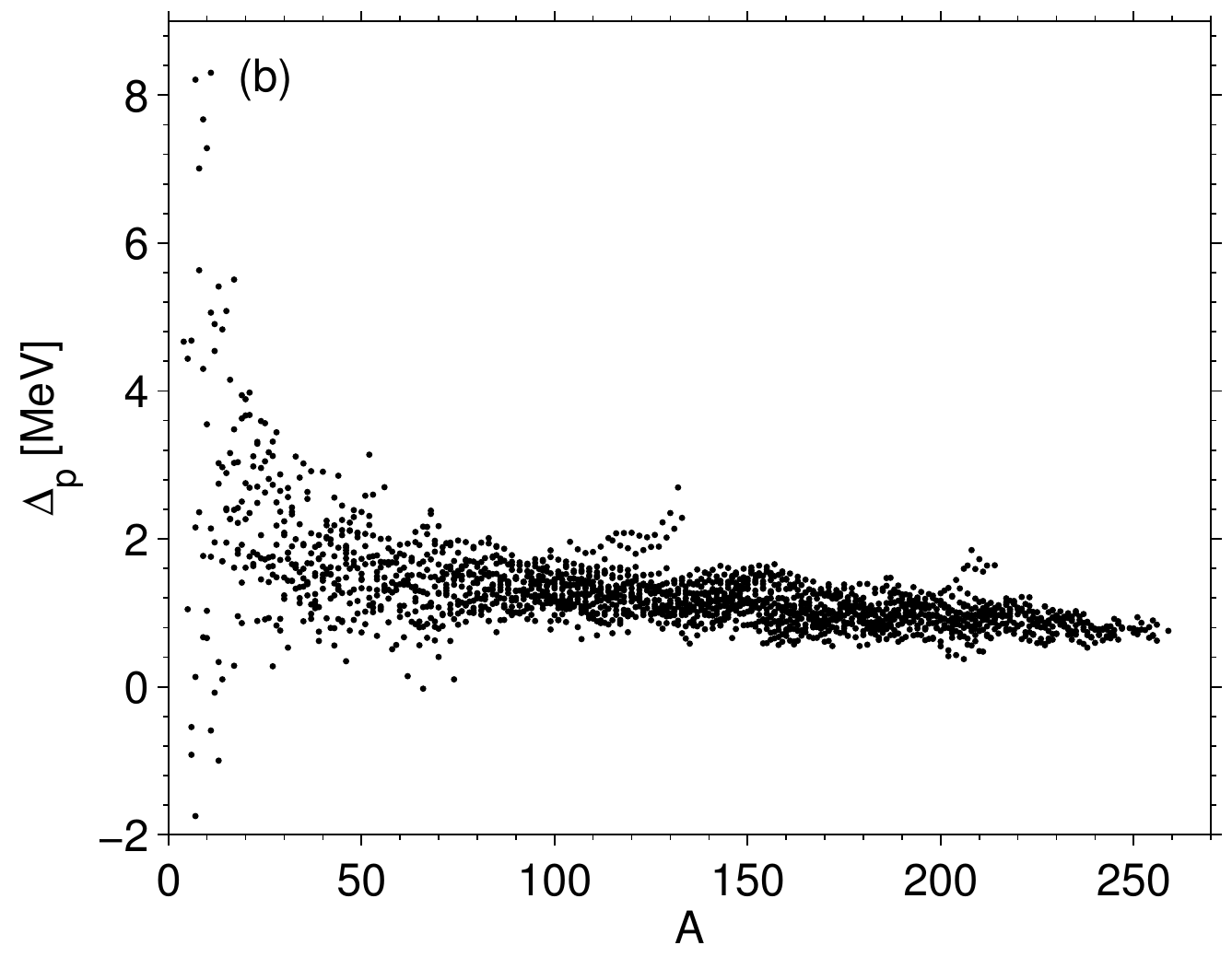} \label{pure p}
 }
 
\caption{The odd-even staggering $\Delta_n$ and $\Delta_p$ as a function of $A$ evaluated using Eqs.~(\ref{Dn}) and (\ref{Dp}). \label{pure}}
\end{figure}

Before any description of general tendencies is attempted the
odd-even effects in isolation are presented. Applying the fundamental
$\Delta_n$ and $\Delta_p$ relations from Eqs.~(\ref{Dn}) and
(\ref{Dp}) results in Figs.~\ref{pure}\subref{pure n} and
\ref{pure}\subref{pure p} respectively. These figures very directly
show the odd-even effects in almost complete isolation. All available
nuclei have been included as these figures' main purpose is to offer a
general impression of the effects. Later, when making more
quantitative examinations, some nuclei influenced by unaccountable
effects are excluded.

The global behavior seen in the figures is well known. Included in both figures are
just over 2100 nuclei, and their conformity immediately suggests a
deeper lying structure. Initially, the decline in energy could suggest
a power law dependence on A, but there is a considerable scatter and
also clear substructures, most noticeably for $\Delta_n$ with $ A >
100$ as we shall see later. These structures will be examined in
detail in the following sections.

It is also important to note the similarities between
Fig.~\ref{pure}\subref{pure n} and Fig.~\ref{pure}\subref{pure
  p}. Both the scale and the general structure is essentially
identical.  This similarity between the results for neutron and proton
mass relations occurs in all following examinations. To avoid tedious
repetitions, figures displaying proton mass relations will not be
included.

Although a few very light isotopes ($A<10$) have rather large values,
the general scale is almost constant around $\sim 0.5-1.5 \,
\si{\mega\electronvolt}$. This change in structure from the light to
the heavy nuclei has led to the suggestion of more sophisticated
models in place of the simplest power law dependence used
traditionally. Friedman and Bertsch suggested \cite{ber09} a two term
expression given by $\Delta = c_1 + c_2/A$, which provides a more than
reasonable global description of the odd-even staggering. This
description will be examined more closely in Sec.~\ref{Sec. Glo}.

\subsection{General structure of the odd-even effect \label{Sec. E-O}}

\begin{figure*}
\centering

\subfloat{
   \includegraphics[width=1\textwidth]{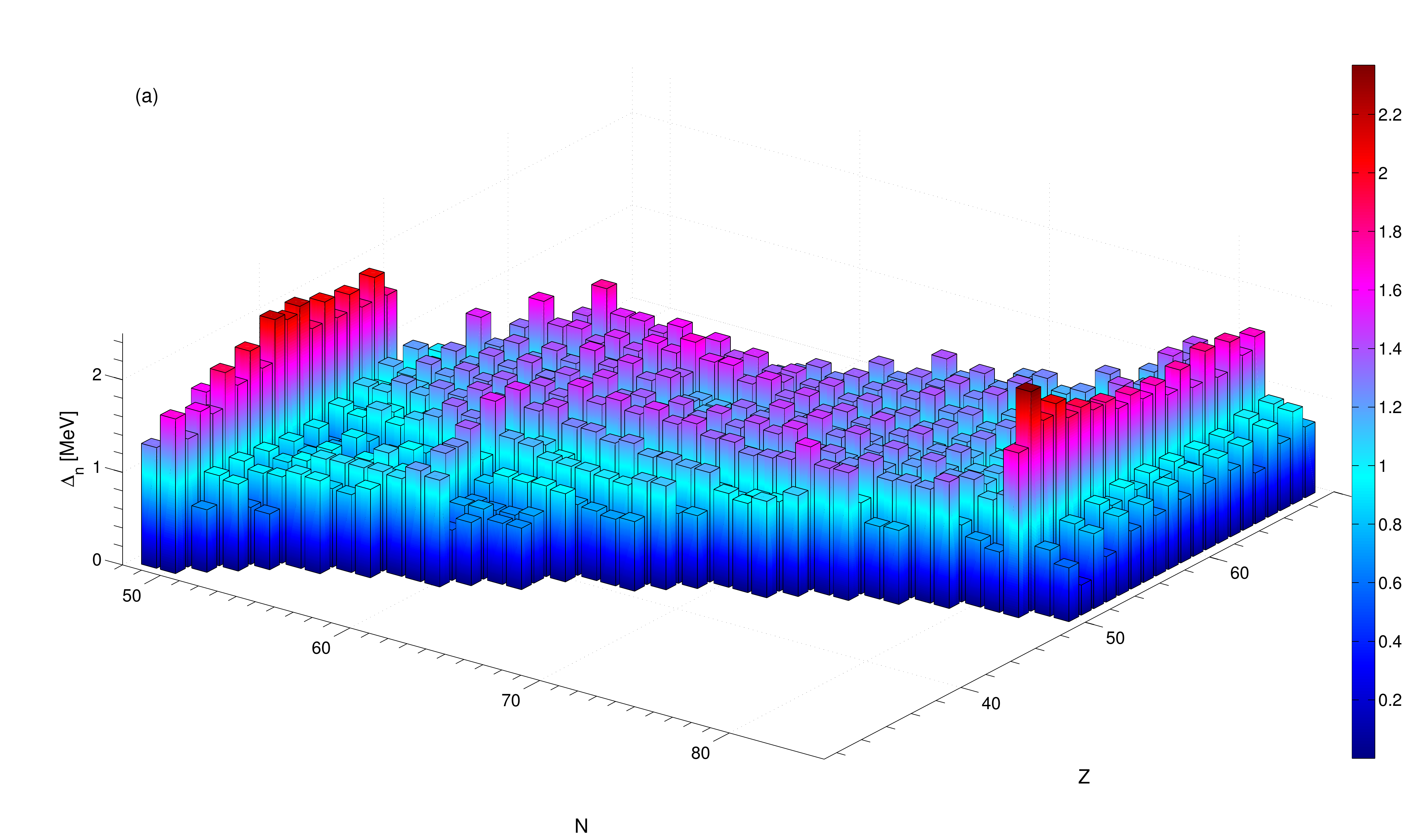} \label{3D light}
 }

\subfloat{
   \includegraphics[width=1\textwidth]{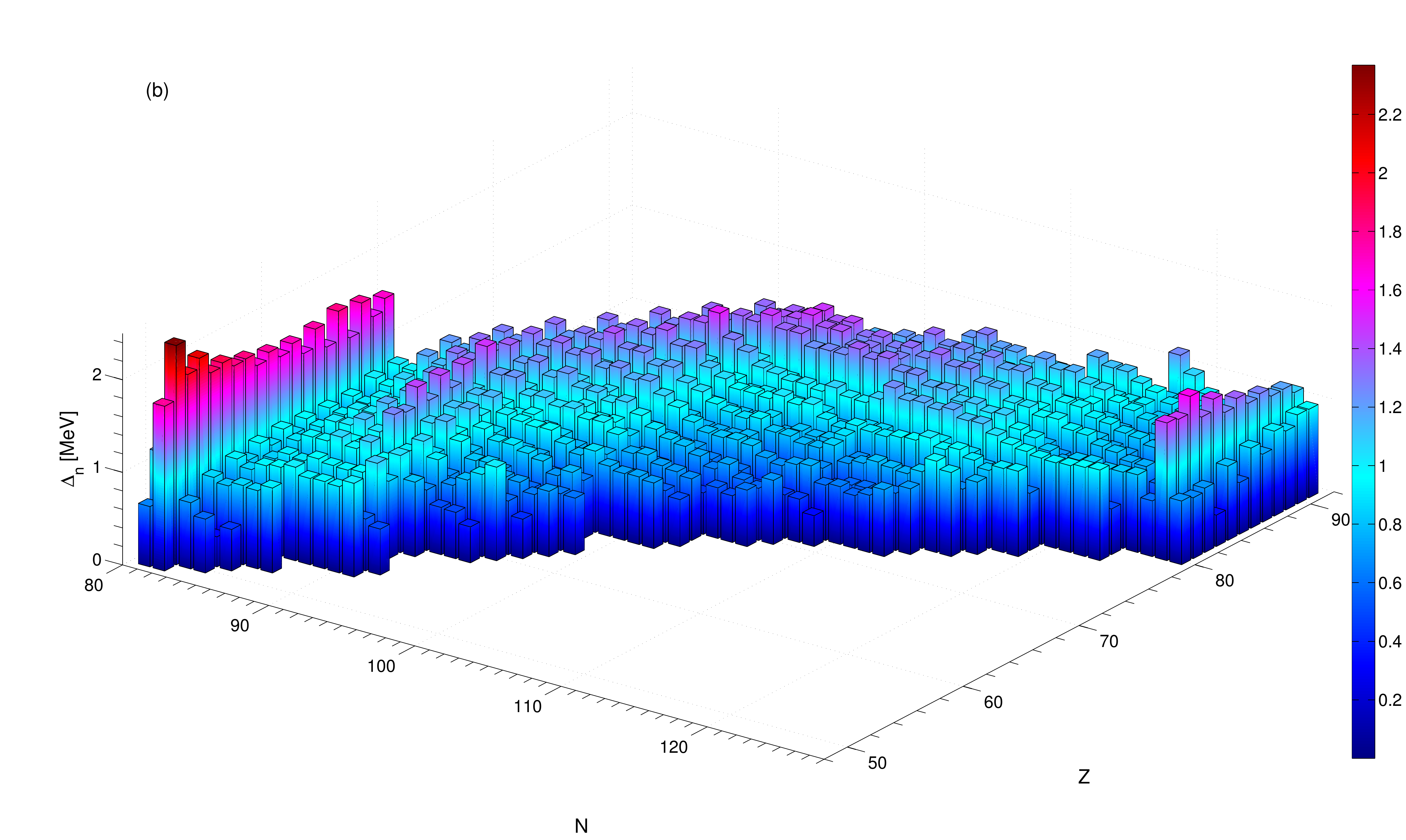} \label{3D heavy}
 }
 
\caption{(Color online) The $(N,Z)$ dependence of $\Delta_n$. Figure (a) shows $49<N<84$ and $30<Z<69$, and figure (b) shows $81<N<127$ and $49<Z<91$. \label{3D}}
\end{figure*}

A more detailed view of the odd-even effects could be useful when
attempting to identify general trends.  Trying to describe the effects
as functions of only one variable is an unfounded restriction. The
possibilities are limited, even if expressing $\Delta_{n}$ and
$\Delta_{p}$ as a function of $A$ or $N-Z$. In Fig.~\ref{3D}
$\Delta_n$ is shown as a function of both $N$ and $Z$, which provides
a detailed look at the actual structure of the staggering effect in
the table of nuclides.

The purpose of these three dimensional figures is to study the
substructures visible in Fig.~\ref{pure} more closely, and to
determine which kind of odd-even effects are immediately visible.
Inspired by Fig.~\ref{pure} the three dimensional figures are limited
to the heavier isotopes. The effect of shells are clearly seen around
magic numbers, but as expected they are very localized.
Fig.~\ref{3D}\subref{3D light} shows the isotopes between the neutron
shells at $N = 50$ and $N = 82$ and Figure \ref{3D}\subref{3D heavy}
the isotopes between the neutron shells at $N = 82$ and $N = 126$. The
proton number has an obvious effect on $\Delta_n$: when $Z$ is odd
$\Delta_n$ is significantly lower compared to the neighboring even-$Z$
nuclei. In both figures another, albeit smaller, odd-even effect is
also seen when changing neutron number. When $N$ is even $\Delta_n$ is
slightly smaller for the neighboring odd $N$ nuclei.

In addition to the neutron shells a smaller effect at the magic proton
numbers $Z = 50$ and $Z=82$ is also seen. Given that a correlation
between $\Delta_n$ and proton number has already been established,
this result is only somewhat surprising. The effect is also less
sharply defined compared to the effect of neutron shells.

Generally, $\Delta_n$ has a clear tendency to decrease away from $N =
Z$, and the tendency is more pronounced for heavier nuclei. This could
indicate a neutron excess dependency, which will be examined in detail in
Sec.~\ref{Sec. Gen}.

\subsection{Extent of shell effects \label{Sec. Shell}}

As seen in Fig.~\ref{3D} the effect of nuclear shells around magic
numbers is very distinct. Although it appears to be very localized, a
more precise estimation of the actual extent would be useful. The
extent can be estimated by applying the fundamental relations from
Eqs.~(\ref{Dn}) and (\ref{Dp}) to a quantitative expression of the
shell effect. To this end Dieperink and Van Isacker's \cite{Diep}
general expression is used.
\begin{align}
E_{shell}(N,Z) =& ( -1.39 S_2 + 0.020 (S_2)^2  \notag \\ 
&+ 0.003 S_3 + 0.075 S_{np} ) \, \si{\mega\electronvolt}, \label{Eq. Shell}
\end{align}
where
\begin{align}
S_2 
&= \frac{n_v \bar{n}_v}{D_n} + \frac{z_v \bar{z}_v}{D_z}, \notag \\
S_3 
&= \frac{n_v \bar{n}_v ( n_v - \bar{n}_v)}{D_n} + \frac{z_v \bar{z}_v (z_v - \bar{z}_v)}{D_z}, \notag \\
S_{np}
&= \frac{n_v \bar{n}_v z_v \bar{z}_v}{D_n D_z}.
\end{align}
Here $n_v$ and $z_v$ are the number of valens nucleons or holes, and
$D_{n,z}$ is the degeneracy of the shell. Finally, $\bar{n}_v \equiv
D_n - n_v$ and $\bar{z}_v \equiv D_z - z_v$. The magic numbers used by
Dieperink and Van Isacker are 2, 8, 14, 28, 50, 82, 126, 184.

\begin{figure}
\centering
\includegraphics[width=1\columnwidth]{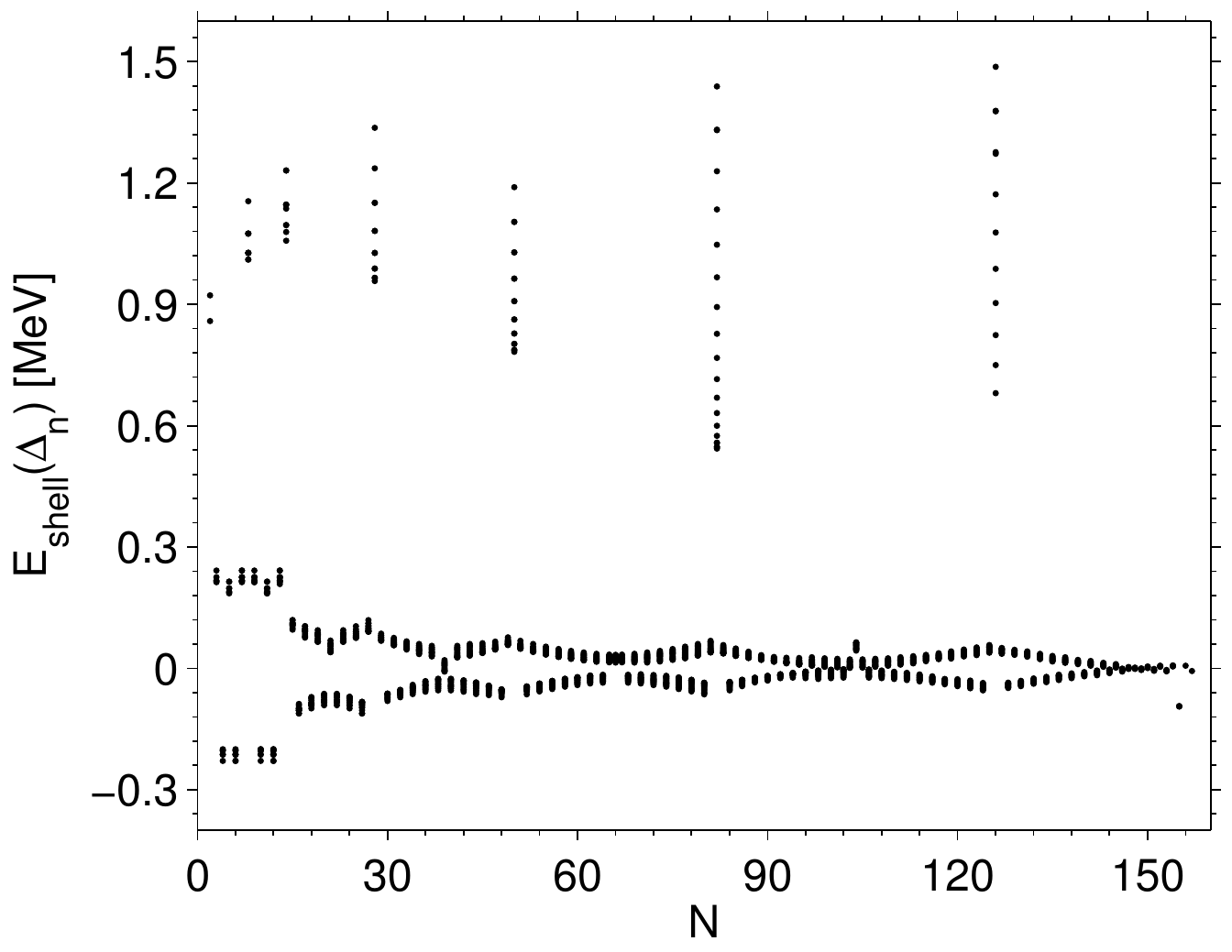} 
\caption{The extent of shell effects. Values calculated using Dieperink and Van Isacker's expression Eq.~(\ref{Eq. Shell}) are combined as in Eq.~(\ref{Q}) using $Q(1,0)$ to estimate the remaning shell contribution for $\Delta_n$. \label{Fig. Shell}}
\end{figure}

The result is presented in Fig.~\ref{Fig. Shell}, and as expected the
effect is extremely localized. A contribution on the scale of $1 \,
\si{\mega\electronvolt}$ around magic numbers seems reasonable when
compared to Fig.~\ref{3D}, but otherwise the effect is less than $\sim
0.1\si{\mega\electronvolt}$, and essentially negligible. The same
localized result  with respect to $Z$ is found when using
Eq.~(\ref{Dp}) instead of Eq.~(\ref{Dn}).

Based on these results the exclusion of mass relation evaluations
which includes magic numbers should be justified. Unless otherwise
stated no corrections are made for the shell effects, the relevant
nuclei are merely excluded from the calculations.

Alternatively, the nuclei, which deviates because of shell effects
could be determined by using a mass relation like Eq.~(\ref{Q}) with
$n=2$ and $z=2$. Then the odd-even effects would be canceled, and the
shell effects would be left in relative isolation. This is done
successfully in detail in \cite{HJR} with a four nucleus mass
relation.



\section{Detailed segmented analysis \label{Sec. Gen}}

Section \ref{Sec. Iso.} looks at a possible neutron-excess dependency
for $\Delta_n$ and $\Delta_p$. The nuclei are separated into groups
according to even and odd neutron and proton numbers. In addition, the
nuclei are divided into regions defined by shells in a similar manner
as in ref. \cite{Jan07} in order to see possible related structure
changes.  In Sec.~\ref{Sec. Glo} the separated nuclei are combined
into a globally valid model of the odd-even effects. This is in
Sec.~\ref{Sec. Com} compared to a global, two term description, with
an $A^{-1}$ dependence.

\subsection{Neutron-excess dependency \label{Sec. Iso.}}

As indicated by Figs.~\ref{3D}\subref{3D heavy} and \ref{3D}\subref{3D
  light} the general nature of the staggering effects seems to change
around $Z=50$. In order to examine this possibility closer the
following results are divided into areas defined by shells.

The structures in question given by Eqs.~(\ref{Dn}) and (\ref{Dp})
involves an odd number of nuclei as seen in Fig.~\ref{Struc.}. Either
a change in proton number or neutron number should then result in a
staggering effect, as a result of Pauli's principle. To thoroughly
explore both possible staggering effects we consider separately nuclei
with $(N,Z)$ being even-even, even-odd, odd-even, and odd-odd.

The isospin projection dependency employed here is, as in \cite{Vog84,Aksel}, a
scaled ($A^{1/3}$)  quadratic neutron-excess  dependency
\begin{align}
A^{1/3} \Delta_{n,p} = a \left( \frac{N - Z}{A} \right)^2 + b, \label{Eq. Iso}
\end{align}
where $a$ and $b$ are constants.

\begin{figure}
\centering
\includegraphics[width=1\columnwidth]{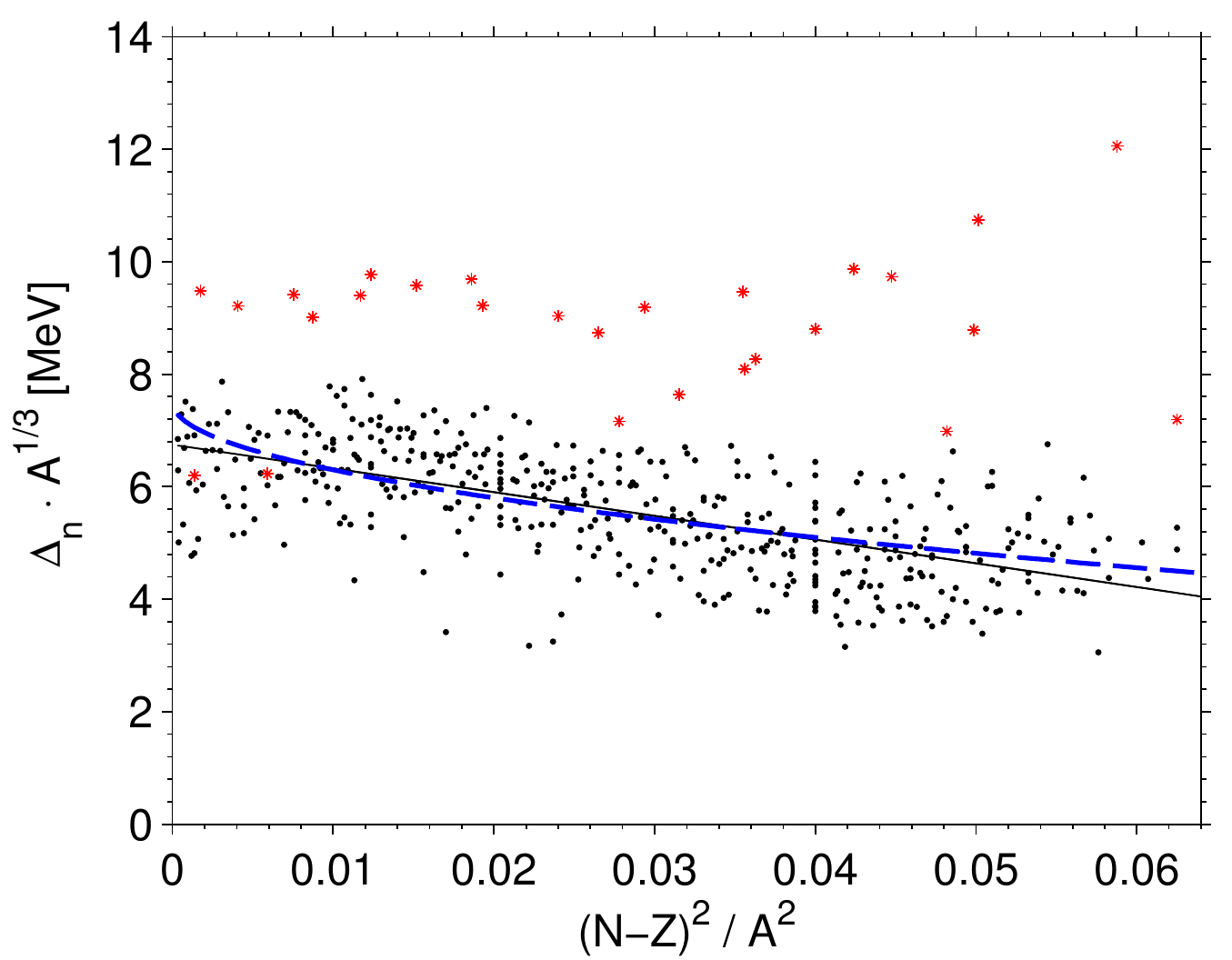}
\caption{(Color online) The $\left( \frac{N-Z}{A} \right)^2$ dependency for all even-even nuclei with $A>50$. The line is the best possible fit based on Eq.~(\ref{Eq. Iso}). The nuclei indicated in red (fat and full) are influenced either by the Wigner effect or by shell effects and are not included in the calculations. The dashed, blue line is the best result when using a model with a linear dependence on the absolute value of neutron excess, $\left\lvert \frac{N-Z}{A} \right\rvert$. \label{Iso e-e tot}}
\end{figure}

A linear neutron excess dependence was suggested in connection with
the Duflo-Zuker mass formula \cite{Duflo}. The corresponding global
parametrization with a single parameter was in \cite{Wang} added to a
liquid drop formula to describe the odd-even staggering. As is seen
in Fig.~\ref{Iso e-e tot} the data do not clearly distinguish the
different functional forms of the neutron excess dependence. The
dashed, blue line is the best fit with a linear (absolute value)
neutron excess dependence, as used in the Duflo-Zuker mass formula
\cite{Duflo}. The difference between the curves is minute compared to
the scatter of the points.  When evaluating the root mean squared
error of $\Delta_{n,p}$ a difference of $0.01 \si{\mega\electronvolt}$ out of a value of $0.16 \si{\mega\electronvolt}$
is found for the two functional forms. Comparing fits for a series of
different regions in the isotopic map we find differences amounting
only by at most $0.01 \si{\mega\electronvolt}$ for all cases shown in
Table \ref{Tab. Iso} for the quadratic relation.
 
We shall focus on nuclei with $A>50$, where the one-parameter, linear
neutron excess dependence is less suited for extrapolation into
unknown mass regions. This is mainly due to the linear form which
would increase $\Delta_p$ away from stability for the proton rich
nuclei with $N>Z$. The global value of the single parameter would
produce less accurate results for any specific region of nuclei. This
global versus local parametrizations will be demonstrated in Table
\ref{Tab. Iso} in connection with descriptions based on
Eq.~(\ref{Eq. Iso}). Thus, we choose to use the quadratic structure
as it is naturally obtained, when expanding with respect to nucleon
number and neutron excess as in droplet models \cite{Myers}. We shall
briefly return to this question in in Sect.~\ref{Sec. Com}.

We have also explored whether other functional forms could reproduce
the dependency seen in the two-dimensional plots. In particular, fits
have been made replacing the $(N-Z)^2/A^2$ term in Eq.~(\ref{Eq. Iso}) with $P =
n_vz_v/(n_z+z_v)$ that has been used to trace the transition to
collective behaviour \cite{Casten}.  Such transitions depend on the
distance from closed shells precisely as the shell effects which as
well might be a plausible reasons for the decreasing $\Delta_{n,p}$
values from magic numbers towards the middle of the shells.  The
resulting fits for $50<Z$ are better than with constant terms, but not
as good as with the $(N-Z)^2$ dependence. Furthermore, the sign of the
coefficient in front of $P$ differs for different regions whereas the
isospin term has a more consistent behaviour. Fits have also been
performed replacing the $A^{1/3}$ in Eq.~(\ref{Eq. Iso}) with
$A^{1/2}$ and $A$. Rather similar overall quality of fits were
obtained, but with a slight preference for $A^{1/3}$ over $A^{1/2}$,
and somewhat better than the $1/A$ dependence.

In all following calculations nuclei influenced by shell effects or
the Wigner effect \cite{Myers} are excluded. They are, however, indicated
in red (fat and full) in the figures. All light nuclei with $A<50$ are
also excluded.

Figure \ref{Iso e-e tot} shows the result of applying
Eq.~(\ref{Eq. Iso}) to all relevant even-even nuclei for $A^{1/3}
\Delta_n$. As expected the excluded nuclei indicated in red (fat and
full) deviate significantly from the observed tendencies. These
tendencies are otherwise reasonable well described by
Eq.~(\ref{Eq. Iso}), but upon closer inspection Fig.~\ref{Iso e-e tot}
appears to be a combination of two straight lines; an almost constant
line around $\sim 6 \, \si{\mega\electronvolt}$, and a slightly
decreasing line superimposed on the first. 

\begin{figure}
\centering
\subfloat{
   \includegraphics[width=1\columnwidth]{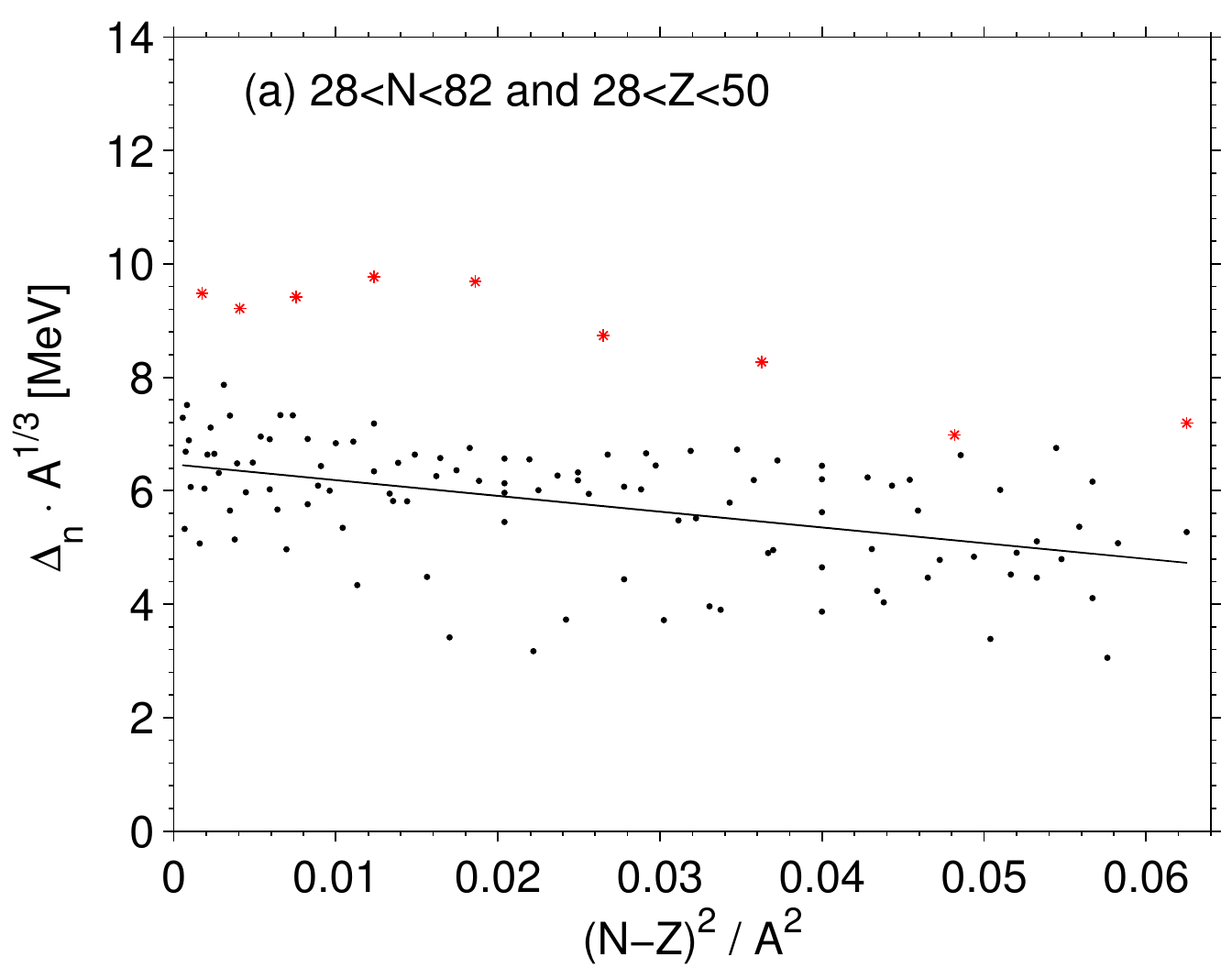} \label{e-e low}
}

\subfloat{
   \includegraphics[width=1\columnwidth]{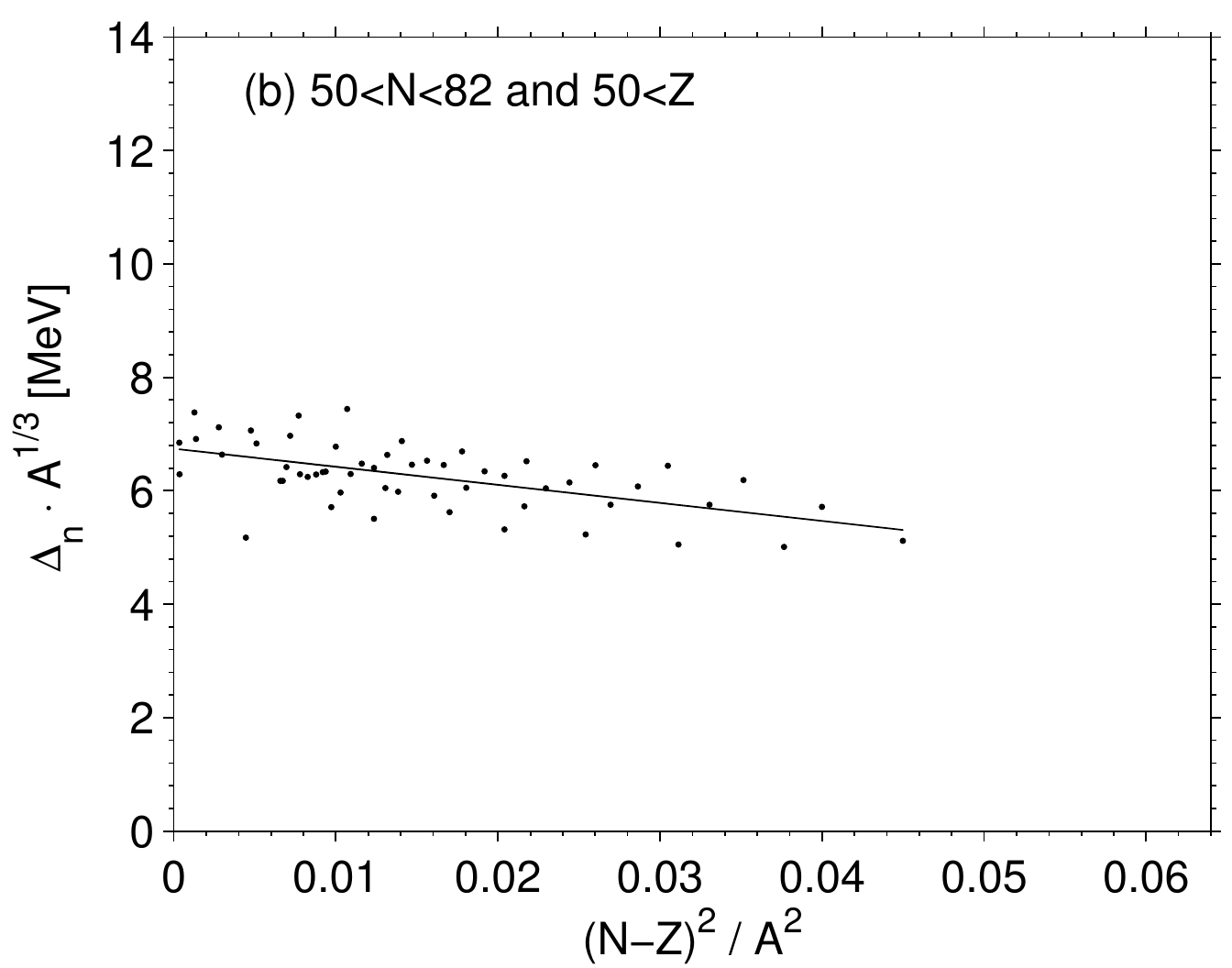} \label{e-e mid}
}

\subfloat{
   \includegraphics[width=1\columnwidth]{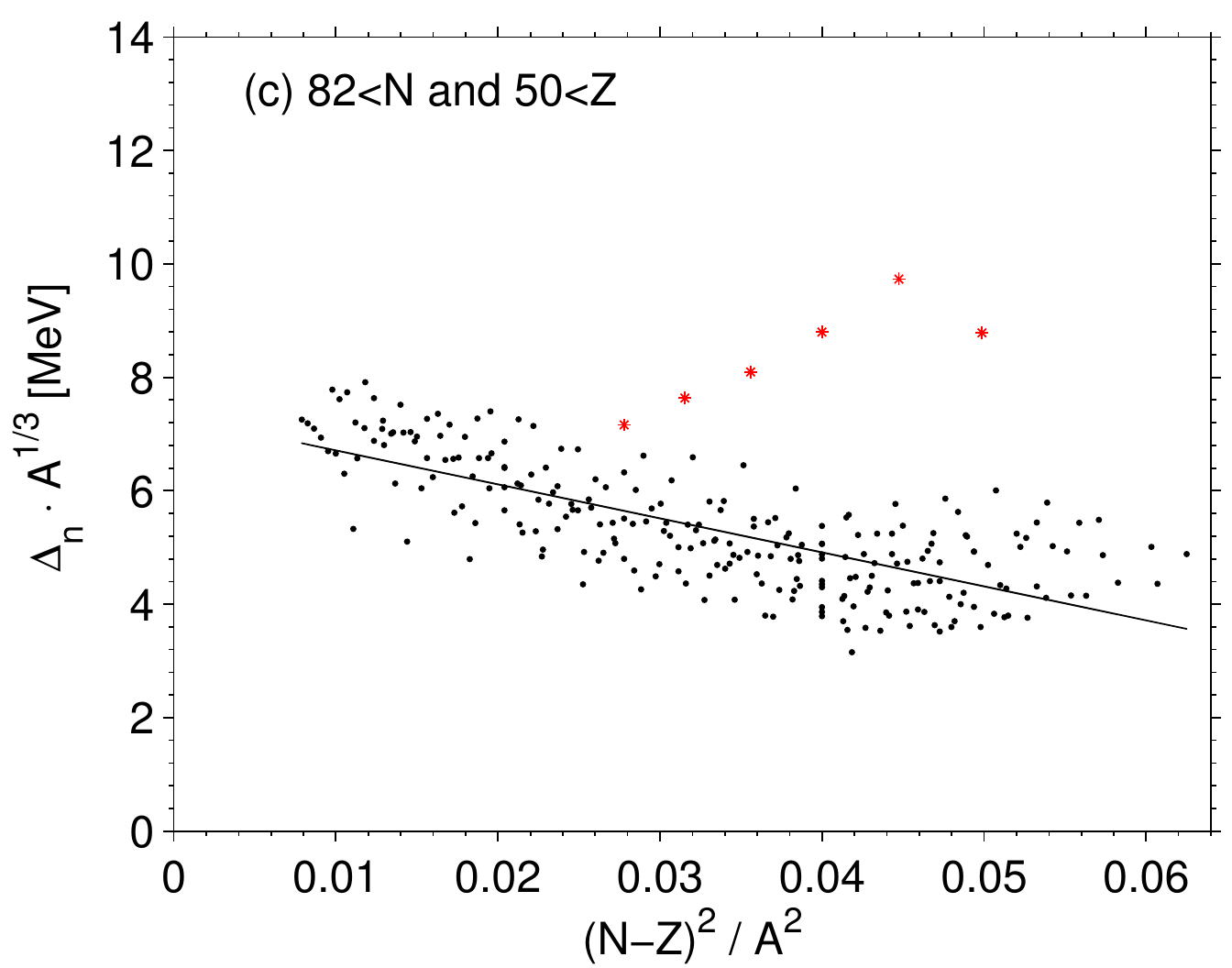} \label{e-e high}
}
\caption{The $\left( \frac{N-Z}{A} \right)^2$ dependency of even-even nuclei divided into regions defined by major nuclear shells as given in the figures. \label{Iso e-e}}
\end{figure}

This is examined more closely by separating the nuclei in three 
regions. The data for even-even nuclei are shown in Fig.~\ref{Iso e-e}
and the parameters of the fitted lines are given in Table
\ref{Tab. Iso}. Figure \ref{Iso e-e}\subref{e-e low} shows the result for the region
given by $28<N<82$, and $28<Z<50$. Likewise, Fig.~\ref{Iso
  e-e}\subref{e-e mid} is for $50<N<82$, and $50<Z$, and \ref{Iso
  e-e}\subref{e-e high} is for $82<N$, and $50<Z$. 
It is clear that the neutron-excess  dependency is less pronounced
for lighter nuclei. In Fig.~\ref{Iso e-e}\subref{e-e low} the results
are almost constant, when considering the scattering, despite the fact
that the $a$-coefficient indicates a small decrease. In Fig.~\ref{Iso
  e-e}\subref{e-e mid}, and in particular in Fig.~\ref{Iso
  e-e}\subref{e-e high} there is an unmistakable decrease as a
function of isospin.

This change around $Z=50$ occurs for both $\Delta_n$ and $\Delta_p$
with even-even nuclei. The change is less definitive, but still
observable, when considering odd-even nuclei.

\begin{figure}
\centering
\subfloat{
   \includegraphics[width=1\columnwidth]{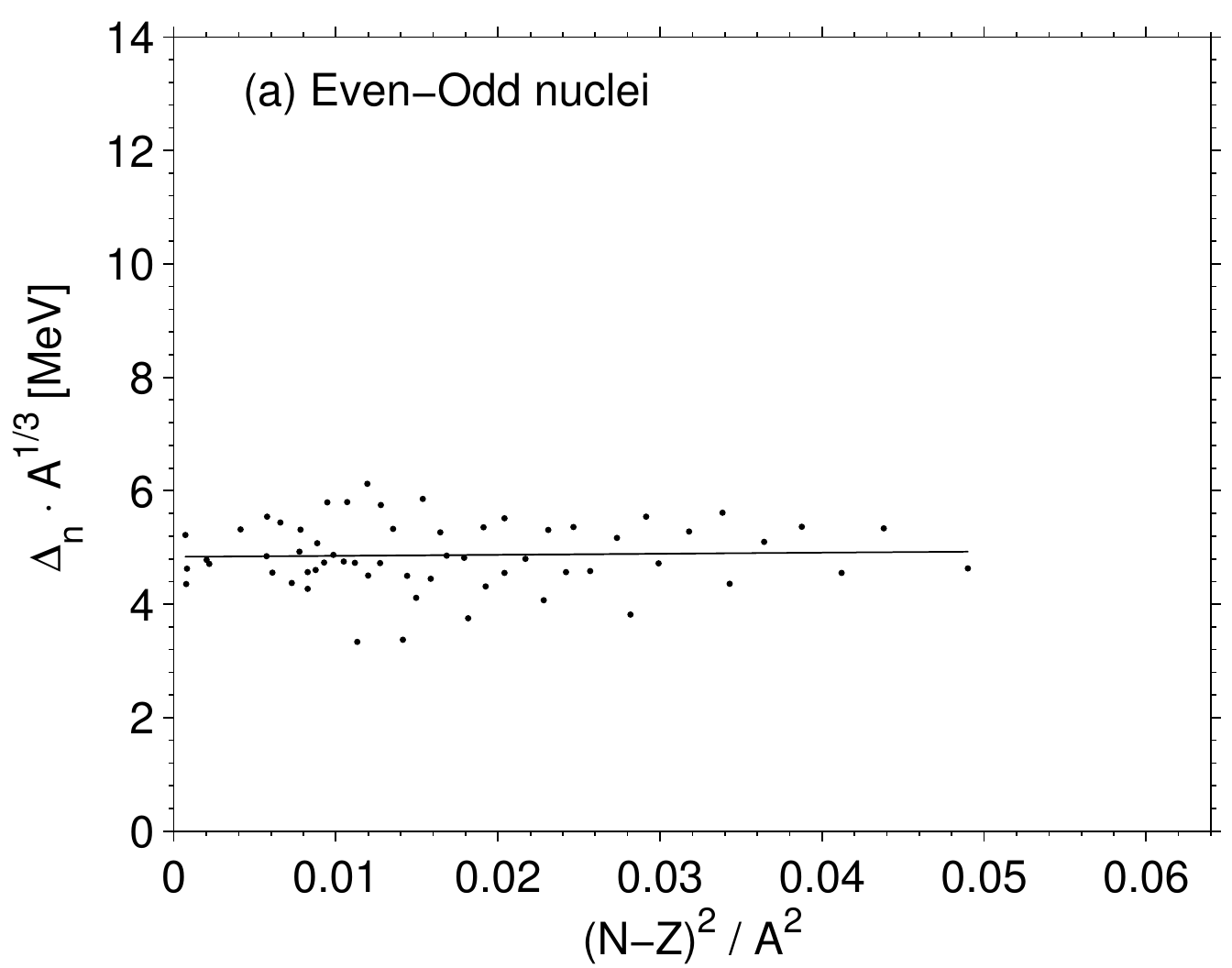} \label{e-o mid}
}

\subfloat{
   \includegraphics[width=1\columnwidth]{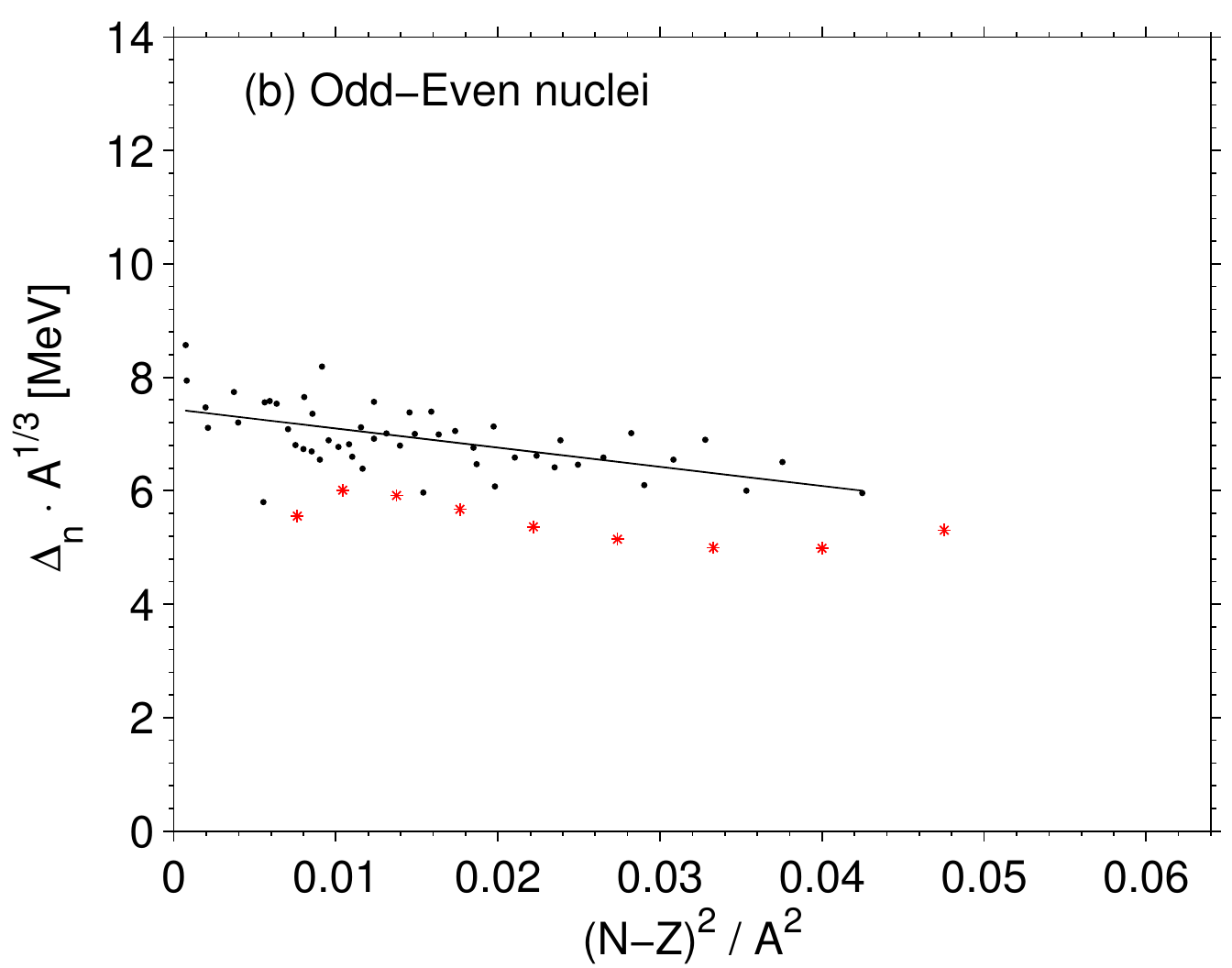} \label{o-e mid}
}

\subfloat{
   \includegraphics[width=1\columnwidth]{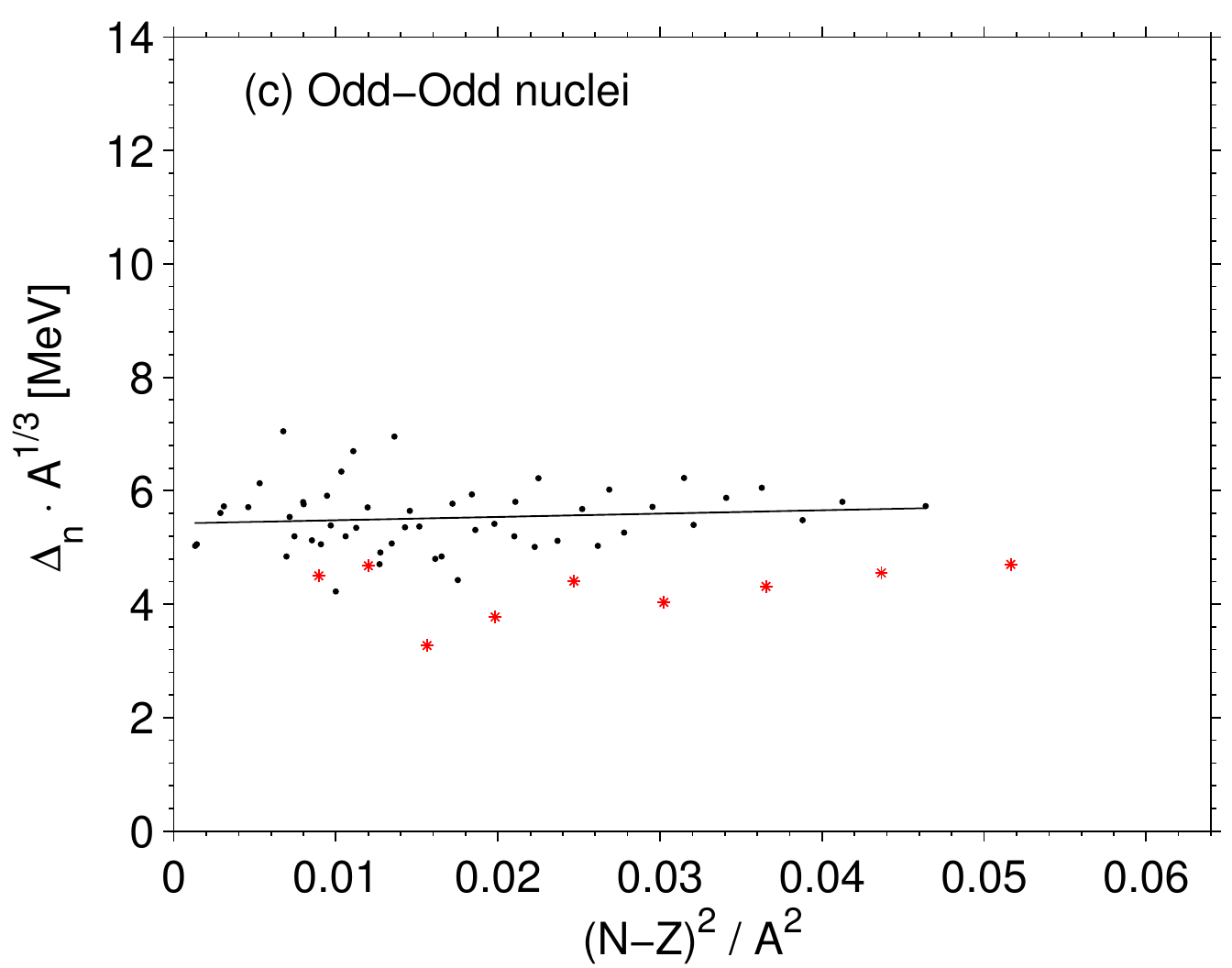} \label{o-o mid}
}

\caption{(Color online) The $\left( \frac{N-Z}{A} \right)^2$  dependency for nuclei with $50<N<82$ and $50<Z$
  divided according to whether $(N,Z)$ is even-odd, odd-even or odd-odd. The even-even nuclei are included in Fig.~\ref{Iso e-e}. \label{Iso mid}}
\end{figure}

The effect of separating the nuclei according to even-odd, odd-even,
and odd-odd is seen in Fig.~\ref{Iso mid}, where the region with
$50<N<82$ and $50<Z$ is presented for the three configurations. Figure
\ref{Iso mid}\subref{e-o mid} shows the result for even-odd, Fig.~\ref{Iso
  mid}\subref{o-e mid} for odd-even, and Fig.~\ref{Iso mid}\subref{o-o mid}
for odd-odd. The even-even nuclei are shown in Fig.~\ref{Iso
  e-e}\subref{e-e mid}.

The results for even-odd and odd-odd when accounting for scattering
are constant, whereas even-even and odd-even very clearly decrease.
This superficially indicates that neutrons and protons behave
differently.  However, their numbers and the valence shells are
different as well.  For $\Delta_p$ in the region where $50<N<82$ and
$50<Z$ a ``symmetric'' result is found: even-even and even-odd have a
clear $N-Z$ dependence, while odd-odd and odd-even are
constant. However, all (even-even, even-odd, odd-even, odd-odd for
both $\Delta_n$ and $\Delta_p$) show a clear $N-Z$ dependency for the
heavier nuclei where $82<N$ and $50<Z$.

The change in neutron-excess dependency at $Z=50$ for the staggering effect
of one type of nucleon is clearly connected to the other nucleon
type.

The parameters of the best linear fits in $(N-Z)^2/A^2$ are collected
in Table \ref{Tab. Iso} that includes the results for even-even,
even-odd, odd-even, and odd-odd nuclei for both $\Delta_n$ and
$\Delta_p$. The overall results are presented as well as results for
three regions in the nuclear chart. The structure changes seen at
$Z=50$ are even more pronounced for $\Delta_p$. Also included in Table
\ref{Tab. Iso} is the root mean squared error (RMSE) for all four
quantities, $A^{1/3}\Delta_{n,p}$ and $\Delta_{n,p}$.

At first glance the fitted parameters seems to reflect rather
different dependencies, simply because of the large variation of the
parameters.  However, all the constant shifts, $b$, are between $5$
and $8$~MeV each with uncertainties of about $0.2$~MeV. The slope parameter,
$a$, has a much larger range of variation but also determined with
much larger uncertainty.  In the region where $28<N<82$ and $28<Z<50$,
the neutron shell at $N=50$ does not signal change of behavior and
therefore the full region is always included.  The $a$-parameter is
sometimes consistent with zero or very small reflecting the flat
behavior discussed in connection with the Figs.~\ref{Iso e-e} and
~\ref{Iso mid}.

It is highly significant that the overall uncertainty of the
$\Delta_{n,p}$-values are very small.  The RMSE is always
(significantly) less than $0.20 \, \si{\mega\electronvolt}$
demonstrating that the parametrization reproduce the observed values
very well.  These absolute uncertainties are comparable to the
underlying chaotic component of $0.1-0.2$~MeV in nuclear masses
\cite{mol06}. This can usually be considered as a lower limit for
systematic reproduction of nuclear binding energies.  In addition,
this suggests that the division into individually fitted regions are
unnecessary.  We shall return and explore this avenue in the next
subsection.

The current data reach out to $(N-Z)^2/A^2$ around 0.05--0.06 and
$\Delta_n$ has for the heaviest region, where $82<N$ and $50<Z$, by
then decreased by a factor two. A naive extrapolation would give zero
odd-even staggering for $(N-Z)^2/A^2$ around 0.10--0.12. This value is
obtained for $N \simeq 2Z$ which is not too far from estimates of the
neutron dripline position. There is no physical basis for
extrapolating this far, but let us briefly discuss the implications
this has.

In the traditional interpretation zero odd-even staggering implies that the cost, $d$, of
one lifted particle at the dripline precisely has to be compensated
by the pairing gain,  $d \approx 0.5 \Delta^2/d$, where $\Delta$
is the usual pairing gap, that is $d = \Delta/\sqrt{2}$.  However,
looking further into the basic meaning quickly reveal inconsistencies,
because also $\Delta$, as proportional to the odd-even mass difference,
has to vanish.
Speculations about small $d$ due to small binding energy and/or
vicinity to the continuum is not convincing, since close-lying levels
usually produce larger pairing gap and pairing energy gain.
Therefore, first the indication of small odd-even mass difference at
the dripline is based on an extrapolation and therefore not in itself
sufficient evidence.  Second, we emphasize that the reason for
vanishing gaps is due to coupling between neutrons and protons simply
because the decrease is as function of the neutron excess.

\begin{table}
\begin{tabular}{l c c c c r r r  c  r} 
\colrule
Type               & Nuclei & \multicolumn{2}{c}{Region limits} & & \multicolumn{1}{c}{a} & \multicolumn{1}{c}{b} & RMSE   & RMSE $(\Delta )$   \\                                                   
\cline{3-4} \cline{6-9}
    &  (N,Z)  & $N$ & $Z$ && \multicolumn{4}{c}{$\left[ \si{\mega\electronvolt} \right]$}     \\
\colrule
$A^{1/3} \Delta_n$ & (e, e)   & All      & All      & & $-42(5)$   &   $6.7(2)$  & $0.81$ & $0.16$  \\
$A^{1/3} \Delta_n$ & (o, e)   & All      & All      & & $-58(6)$   &   $7.6(2)$  & $0.96$ & $0.19$  \\
$A^{1/3} \Delta_n$ & (e, o)   & All      & All      & & $-30(5)$   &   $5.1(2)$  & $0.80$ & $0.16$ \\
$A^{1/3} \Delta_n$ & (o, o)   & All      & All      & & $-44(6)$   &   $5.9(2)$  & $0.91$ & $0.18$  \\
$A^{1/3} \Delta_n$ & (e, e)   & $28, 82$ & $28, 50$ & & $-28(10)$  &   $6.5(3)$  & $0.93$ & $0.20$  \\
$A^{1/3} \Delta_n$ & (o, e)   & $28, 82$ & $28, 50$ & & $-41(12)$  &   $7.4(4)$  & $1.06$ & $0.23$  \\
$A^{1/3} \Delta_n$ & (e, o)   & $28, 82$ & $28, 50$ & & $-15(8)$   &   $4.6(2)$  & $0.76$ & $0.16$  \\
$A^{1/3} \Delta_n$ & (o, o)   & $28, 82$ & $28, 50$ & & $-26(9)$   &   $5.4(3)$  & $0.87$ & $0.19$  \\
$A^{1/3} \Delta_n$ & (e, e)   & $50, 82$ & $50, -$  & & $-32(12)$  &   $6.7(2)$  & $0.48$ & $0.09$  \\
$A^{1/3} \Delta_n$ & (o, e)   & $50, 82$ & $50, -$  & & $-34(13)$  &   $7.4(2)$  & $0.47$ & $0.09$  \\
$A^{1/3} \Delta_n$ & (e, o)   & $50, 82$ & $50, -$  & & $2(14)$    &   $4.8(3)$  & $0.59$ & $0.12$  \\
$A^{1/3} \Delta_n$ & (o, o)   & $50, 82$ & $50, -$  & & $6(15)$    &   $5.4(3)$  & $0.57$ & $0.11$  \\
$A^{1/3} \Delta_n$ & (e, e)   & $82, -$  & $50, -$  & & $-60(7)$   &   $7.3(2)$  & $0.71$ & $0.12$  \\
$A^{1/3} \Delta_n$ & (o, e)   & $82, -$  & $50, -$  & & $-69(8)$   &   $7.8(3)$  & $0.91$ & $0.16$  \\
$A^{1/3} \Delta_n$ & (e, o)   & $82, -$  & $50, -$  & & $-50(8)$   &   $5.8(3)$  & $0.74$ & $0.13$  \\
$A^{1/3} \Delta_n$ & (o, o)   & $82, -$  & $50, -$  & & $-64(9)$   &   $6.5(3)$  & $0.86$ & $0.15$  \\
$A^{1/3} \Delta_p$ & (e, e)   & All      & All      & & $-27(5)$   &   $6.9(1)$  & $0.73$ & $0.15$  \\
$A^{1/3} \Delta_p$ & (o, e)   & All      & All      & & $-13(5)$   &   $5.1(2)$  & $0.76$ & $0.16$  \\
$A^{1/3} \Delta_p$ & (e, o)   & All      & All      & & $-43(6)$   &   $7.9(2)$  & $0.86$ & $0.17$  \\
$A^{1/3} \Delta_p$ & (o, o)   & All      & All      & & $-26(6)$   &   $6.1(2)$  & $0.85$ & $0.17$  \\
$A^{1/3} \Delta_p$ & (e, e)   & $28, 82$ & $28, 50$ & & $-6(7)$    &   $6.5(2)$  & $0.62$ & $0.14$  \\
$A^{1/3} \Delta_p$ & (o, e)   & $28, 82$ & $28, 50$ & & $4(8)$     &   $4.6(2)$  & $0.72$ & $0.16$  \\
$A^{1/3} \Delta_p$ & (e, o)   & $28, 82$ & $28, 50$ & & $-16(9)$   &   $7.5(3)$  & $0.74$ & $0.16$  \\
$A^{1/3} \Delta_p$ & (o, o)   & $28, 82$ & $28, 50$ & & $-2(9)$    &   $5.6(3)$  & $0.78$ & $0.17$  \\
$A^{1/3} \Delta_p$ & (e, e)   & $50, 82$ & $50, -$  & & $-43(16)$  &   $7.2(3)$  & $0.58$ & $0.11$  \\
$A^{1/3} \Delta_p$ & (o, e)   & $50, 82$ & $50, -$  & & $-13(18)$  &   $5.3(4)$  & $0.74$ & $0.15$  \\
$A^{1/3} \Delta_p$ & (e, o)   & $50, 82$ & $50, -$  & & $-39(14)$  &   $7.9(3)$  & $0.46$ & $0.09$  \\
$A^{1/3} \Delta_p$ & (o, o)   & $50, 82$ & $50, -$  & & $-10(19)$  &   $6.0(4)$  & $0.64$ & $0.12$  \\
$A^{1/3} \Delta_p$ & (e, e)   & $82, -$  & $50, -$  & & $-42(7)$   &   $7.4(2)$  & $0.70$ & $0.13$  \\
$A^{1/3} \Delta_p$ & (o, e)   & $82, -$  & $50, -$  & & $-35(7)$   &   $6.0(2)$  & $0.64$ & $0.11$  \\
$A^{1/3} \Delta_p$ & (e, o)   & $82, -$  & $50, -$  & & $-53(9)$   &   $8.0(3)$  & $0.89$ & $0.16$  \\
$A^{1/3} \Delta_p$ & (o, o)   & $82, -$  & $50, -$  & & $-45(9)$   &   $6.6(3)$  & $0.84$ & $0.15$  \\
\colrule
\end{tabular}
\caption{Fit parameters for Eq.~(\ref{Eq. Iso}) divided into regions
  defined by shells, and seperated for even-even, even-odd, odd-even,
  and odd-odd nuclei. RMSE is the root mean squared error of
  $A^{1/3}\Delta_{n,p}$, and RMSE($\Delta$) is the root mean squared
  error of $\Delta_{n,p}$ only. \label{Tab. Iso}}
\end{table}

\begin{figure}
\centering
\includegraphics[width=1\columnwidth]{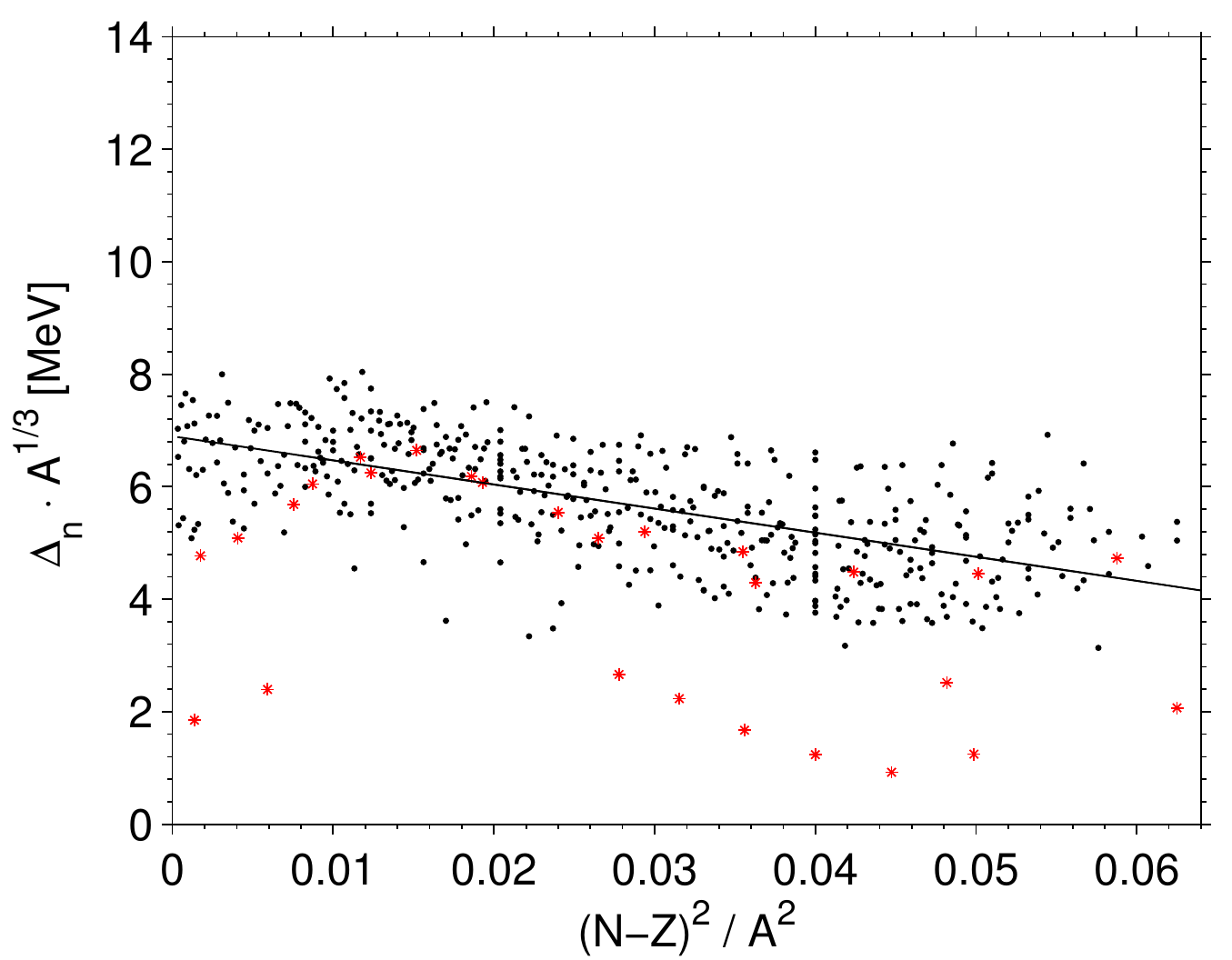}
\caption{(Color online) The $\left( \frac{N-Z}{A}\right)^2$ dependency for all even-even nuclei with $A>50$
  with a shell effect correction given by
  Eq.~(\ref{Eq. Shell}). \label{Shell correc.}}
\end{figure}

The results in Table \ref{Tab. Iso} still include remnants of the
shell effect.  If Eq.~(\ref{Eq. Shell}) is used to correct for shell
effects one obtains Fig.~\ref{Shell correc.} that can be compared to
Fig.~\ref{Iso e-e tot}. The best linear fit is now $A^{1/3} \Delta_n =
-43(5) \si{\mega\electronvolt} \left( \frac{N-Z}{A} \right)^2 +
6.9(2)\si{\mega\electronvolt}$ with $RMSE(\Delta_n) = 0.16
\si{\mega\electronvolt}$. The only nuclei significantly influenced by
the correction are the ones marked in red (fat and full). Based on
Fig.~\ref{Shell correc.}, Eq.~(\ref{Eq. Shell}) seems to
overcompensate for about half of the affected nuclei.  This is most
likely a result of the different liquid drop parameters used here and
in Dieperink and Van Isacker's paper \cite{Diep}.

\subsection{Global descriptions \label{Sec. Glo}}

Although the neutron-excess parametrization when separating into
regions defined by shells allows for a very accurate description of
the odd-even staggering, it also results in a rather cumbersome
model. The scale of the difference when changing between even and odd
isotopes can be inferred from Table \ref{Tab. Iso}. This can be used
to combine some of the separated nuclei.

Changing from odd to even Z with $\Delta_p$ only changes the constant
term $b$ by less than one MeV, and the scaling factors $a$ are mostly overlapping. As a very reasonable approximation
the separation of $\Delta_p$ into even and odd Z can therefore be
ignored. The best linear fits based on Eq.~(\ref{Eq. Iso}) when
neglecting the separation according to even and odd Z are presented in
Table \ref{Tab. even-odd}.

\begin{table}
\begin{tabular}{l l c c c r  r r  c  r} 
\colrule
Type               & Nuclei & \multicolumn{2}{c}{Region limits} & & \multicolumn{1}{c}{a} & \multicolumn{1}{c}{b} & RMSE   & RMSE $(\Delta)$  \\                                                   
\cline{3-4} \cline{6-9}
    &    & $N$ & $Z$ && \multicolumn{4}{c}{$\left[ \si{\mega\electronvolt} \right]$}     \\
\colrule
$A^{1/3} \Delta_n$ & Even Z   & All      & All      & & $-49(4)$   &   $7.1(1)$  & $0.91$ & $0.18$  \\
$A^{1/3} \Delta_n$ & Odd Z    & All      & All      & & $-37(4)$   &   $5.5(1)$  & $0.89$ & $0.18$  \\
$A^{1/3} \Delta_n$ & Even Z   & $28, 82$ & $28, 50$ & & $-33(8)$   &   $6.9(2)$  & $1.04$ & $0.23$  \\
$A^{1/3} \Delta_n$ & Odd Z    & $28, 82$ & $28, 50$ & & $-20(7)$   &   $5.0(2)$  & $0.87$ & $0.19$  \\
$A^{1/3} \Delta_n$ & Even Z   & $50, 82$ & $50, -$  & & $-33(11)$  &   $7.1(2)$  & $0.58$ & $0.11$  \\
$A^{1/3} \Delta_n$ & Odd Z    & $50, 82$ & $50, -$  & & $3(11)$    &   $5.1(2)$  & $0.67$ & $0.13$  \\
$A^{1/3} \Delta_n$ & Even Z   & $82, -$  & $50, -$  & & $-65(5)$   &   $7.6(2)$  & $0.83$ & $0.14$  \\
$A^{1/3} \Delta_n$ & Odd Z    & $82, -$  & $50, -$  & & $-57(6)$   &   $6.2(2)$  & $0.82$ & $0.14$  \\
$A^{1/3} \Delta_p$ & Even N   & All      & All      & & $-35(4)$   &   $7.3(1)$  & $0.86$ & $0.17$  \\
$A^{1/3} \Delta_p$ & Odd N    & All      & All      & & $-19(4)$   &   $5.6(1)$  & $0.86$ & $0.18$  \\
$A^{1/3} \Delta_p$ & Even N   & $28, 82$ & $28, 50$ & & $-10(6)$   &   $7.0(2)$  & $0.79$ & $0.18$  \\
$A^{1/3} \Delta_p$ & Odd N    & $28, 82$ & $28, 50$ & & $1(7)$     &   $5.1(2)$  & $0.85$ & $0.19$  \\
$A^{1/3} \Delta_p$ & Even N   & $50, 82$ & $50, -$  & & $-44(14)$  &   $7.6(3)$  & $0.67$ & $0.13$  \\
$A^{1/3} \Delta_p$ & Odd N    & $50, 82$ & $50, -$  & & $-13(15)$  &   $5.6(3)$  & $0.80$ & $0.16$  \\
$A^{1/3} \Delta_p$ & Even N   & $82, -$  & $50, -$  & & $-48(6)$   &   $7.7(2)$  & $0.82$ & $0.15$  \\
$A^{1/3} \Delta_p$ & Odd N    & $82, -$  & $50, -$  & & $-40(6)$   &   $6.3(2)$  & $0.76$ & $0.13$  \\
\colrule
\end{tabular}
\caption{Similar to Tabel \ref{Tab. Iso}, but the nuclei are separated according to odd and even nucleon number. \label{Tab. even-odd}}
\end{table}

The effect of changing from odd to even N with $\Delta_p$ can most
easily be inferred from Table \ref{Tab. even-odd}. This is by no means
negligible. Instead the effect can be viewed as a constant addition,
and as such it can be accounted for. This would result in an
expression for the odd-even proton staggering of the form
\begin{align}
A^{1/3} \Delta_p = a \left( \frac{N-Z}{A} \right)^2 + b - (1-\pi_n)c/2, \label{Eq. p comb.}
\end{align}
where $a$ and $b$ are the parameters from Eq.~(\ref{Eq. Iso}), and $c$
is the difference between the constants $b$ for even and odd $N$.

The region with $50<N<82$ and $50<Z$ is the most problematic, as
even and odd N have conflicting tendencies. However, any global
description based on a combination of local descriptions must
necessarily be an approximation. The result of displacing odd N
nuclei, and then finding the best linear fit based on
Eq.~(\ref{Eq. Iso}) is given in Tabel \ref{Tab. Tot}. Also included is
the size of the displacement,  c.

The most interesting result is the combined expression for all
nuclei. A noticeable improvement is the reduction in uncertainty for
the constants $a$ and $b$. Though more important is the size of the
root mean squared error, which is comparable to RMSE of $\Delta_p$ for
even-even nuclei in Tabel \ref{Tab. Iso}. This global expression
should then have almost the same overall accuracy as the former
subdivided expressions, while being much more practicable.

\begin{table}
\begin{tabular}{l c c c c r  r r  c  r} 
\colrule
Type               &  \multicolumn{2}{c}{Region limits} & & c & \multicolumn{1}{c}{a} & \multicolumn{1}{c}{b} & RMSE   & RMSE $(\Delta)$   \\                                                   
\cline{2-3} \cline{5-9}
                   &    $N$ & $Z$ && \multicolumn{5}{c}{$\left[ \si{\mega\electronvolt} \right]$}       \\
\colrule
$A^{1/3} \Delta_n$ &  All      & All      & & $1.7$    & $-44(3)$   &   $7.2(1)$  & $0.92$ & $0.18$  \\
$A^{1/3} \Delta_n$ &  $28, 82$ & $28, 50$ & & $1.9$    & $-27(5)$   &   $6.9(2)$  & $0.97$ & $0.21$  \\
$A^{1/3} \Delta_n$ &  $50, 82$ & $50, -$  & & $2.0$    & $-11(9)$   &   $7.0(2)$  & $0.72$ & $0.14$  \\
$A^{1/3} \Delta_n$ &  $82, -$  & $50, -$  & & $1.4$    & $-62(4)$   &   $7.6(1)$  & $0.83$ & $0.15$  \\
$A^{1/3} \Delta_p$ &  All      & All      & & $1.8$    & $-27(3)$   &   $7.3(1)$  & $0.89$ & $0.19$  \\
$A^{1/3} \Delta_p$ &  $28, 82$ & $28, 50$ & & $1.9$    & $-5(5)$    &   $7.0(1)$  & $0.84$ & $0.19$  \\
$A^{1/3} \Delta_p$ &  $50, 82$ & $50, -$  & & $1.9$    & $-26(11)$  &   $7.6(2)$  & $0.80$ & $0.16$  \\
$A^{1/3} \Delta_p$ &  $82, -$  & $50, -$  & & $1.4$    & $-44(4)$   &   $7.7(2)$  & $0.81$ & $0.14$  \\
\colrule
\end{tabular}
\caption{Similar to Tables \ref{Tab. Iso} and \ref{Tab. even-odd}, but without separation according to odd-even configuration. The value of c in Eq.~(\ref{Eq. p comb.}) signifies the displacement of odd nuclei. \label{Tab. Tot}}
\end{table}

A completely analogous combination can be made for $\Delta_n$. Here
the separation into odd and even $N$ is neglected, and the calculated
displacement is from odd to even $Z$. Neglecting the first separation
is a less good approximation for $\Delta_n$ than for $\Delta_p$. The
result is also included in Table \ref{Tab. even-odd}.

The result of displacing odd $Z$ nuclei is presented in Table
\ref{Tab. Tot}, and the value of the displacement is almost identical
to the result for $\Delta_p$. Based on RMSE $\Delta_n$ is as valid as
$\Delta_p$.

It might initially appear as if the $N-Z$ dependency of the proton
staggering is less pronounced than the neutron staggering. As stated
earlier this dependency increases for heavier nuclei. In other words,
the $N-Z$ dependency of the neutron staggering effect is larger, when
neutrons are abundant and analogously for protons. The nuclei examined
generally have a majority of neutrons, and the $N-Z$ dependence of
the neutron staggering is therefore seemingly greater. In the region
where $50<N<82$ and $50<Z$ the $N-Z$  dependence is seen to be greater for
the proton staggering.

To obtain the term which has to be added to the liquid drop model,
Eq.~(\ref{Eq. p comb.}) is combined with Eq.~(\ref{Delta}). However,
we have a term in $\Delta_n$ proportional to $-(1-\pi_p)/2$ and one in
$\Delta_p$ proportional to $-(1-\pi_n)/2$ but otherwise of the same
magnitude, this is, as remarked in section \ref{Sect. theory},
indicative of a neutron-proton pairing term that must be taken out
before the two separate expressions are added. Noting that the
$b$-coefficients for neutrons and protons in Table \ref{Tab. Tot} are
also essentially identical we obtain the following final relations,
for $Z<50$:
\begin{align}
\Delta =& 
 A^{-1/3} \Bigg( \left( \frac{N-Z}{A} \right)^2 (-13 \pi_n -2.5 \pi_p) + 3.4 (\pi_n + \pi_p) \notag \\ 
&+ 1.85 \pi_{np} \Bigg) \si{\mega\electronvolt},   \label{Finaleq_light}
\end{align}
and for $Z>50$
\begin{align}
\Delta =& A^{-1/3} \Bigg( \left( \frac{N-Z}{A} \right)^2 (-28 \pi_n -20.5 \pi_p) + 3.8 (\pi_n + \pi_p) \notag \\ 
&+ 1.55 \pi_{np}  \Bigg) \si{\mega\electronvolt}.  \label{Finaleq_heavy}
\end{align}
The RMSE is still slightly below $0.2$~MeV in these fits.  Note
that the neutron-proton pairing term here is taken to have a
$A^{-1/3}$ mass dependence as the other terms in contrast to earlier
works \cite{Friedman}.  Since all terms depend on $N$ and $Z$ one
should in principle correct for higher-order effects when going from
the mass differences $\Delta_{n,p}$ to the $\Delta$ that should be
included in mass formulas. However, the correction terms are at most
of order $10^{-3}$ and have been neglected. The effects would be
larger for mass relations involving more nuclei, as an example we find
that the values obtained for $\Delta_{n, p}^{(5)}$ are systematically
10--20\% smaller than the ones for $\Delta_{n, p}$.

We could take seriously the observation that for $Z<50$ the $N-Z$
dependence is either zero or very small.  A fit for these nuclei to
Eq.~(\ref{Eq. p comb.}) with $a=0$ gives 
\begin{align}
\Delta = A^{-1/3} \Bigg( 5.9\pi_n+6.6\pi_p +  1.6\pi_{np} \Bigg) \si{\mega\electronvolt},
\end{align}
corresponding to RMSE $= 0.21$ and $0.25$~MeV for $\Delta_p$ and
$\Delta_n$, respectively. As expected these uncertainties are not as good as
obtained by maintaining the $N-Z$ dependent term. They are, however, reasonably close, as well as somewhat simpler.

\subsection{Comparisons \label{Sec. Com}}

To determine the viability of our expression a comparison with
other more established models is useful. The previously mentioned two
term model suggested by Friedman and Bertsch \cite{ber09} yields very
accurate results, and is well suited as a comparison. We shall also
briefly compare to the very similar model presented by Jensen
\textit{et al.} \cite{Aksel}.

The two term expression suggested by Friedman and Bertsch on the basis
of a more detailed physical modeling is
\begin{align}
\Delta_{n,p} = c_1 + c_2/A, \label{Eq. ber}
\end{align}
where $c_1$ and $c_2$ are constants to be determined. In our fits to
this expression we again exclude
light nuclei, and nuclei otherwise influenced by shell effects or the
Wigner effect. The remains of the smooth aspects
are also removed, as in Eqs.~(\ref{Dn}) and (\ref{Dp}). Based on the
tendencies observed in Fig.~\ref{3D}, $\Delta_n$ is separated according
to odd and even $Z$, and vice versa for $\Delta_p$. The relevant nuclei
are also examined both collectively and separated into the two
heaviest regions given by $50<N<82$ and $50<Z$, and
$82<N$ and $50<Z$.

The RMSE for the results is calculated in two ways. First, the error will
be calculated as the RMSE of the nuclei in a given region in relation
to the best local fit based on Eq.~(\ref{Eq. ber}). The result of this
calculation is presented in Table \ref{Tab. Ber}. The intent with the
two term model was never to separate it according to shells, so the
most interesting results are those which covers all regions. Comparing
Tables \ref{Tab. Ber} and \ref{Tab. even-odd} the error for $\Delta_n$
and $\Delta_p$ is seen to be noticeably larger for the two-term
model in Eq.~(\ref{Eq. ber}). For most regions the isospin dependence also gives a smaller
RMSE.

In a second step, the error is calculated as the RMSE of the nuclei in a
region relative to the best global fit again based on
Eq.~(\ref{Eq. ber}). The result is presented in Table
\ref{Tab. Ber2}. The collective result for all the nuclei would be the
same as in Table \ref{Tab. Ber}, and is not included. The RMSE is
inevitably larger than in Table \ref{Tab. Ber}, but it is still very
reasonable.

\begin{table}
\begin{tabular}{l l c c c r  r r  r} 
\colrule
Type               & Nuclei & \multicolumn{2}{c}{Region limits} & & \multicolumn{1}{c}{$c_2$} & \multicolumn{1}{c}{$c_1$} & RMSE     \\                                                   
\cline{3-4} \cline{6-8}
    &    & $N$ & $Z$ && \multicolumn{3}{c}{$\left[ \si{\mega\electronvolt} \right]$}      \\
\colrule
$\Delta_n$ & Even Z   & All      & All      & & $58(5)$    &   $0.7(0)$  & $0.24$  \\
$\Delta_n$ & Odd Z    & All      & All      & & $32(4)$    &   $0.6(0)$  & $0.21$  \\
$\Delta_n$ & Even Z   & $50, 82$ & $50, -$  & & $168(34)$  &   $0.0(3)$  & $0.11$  \\
$\Delta_n$ & Odd Z    & $50, 82$ & $50, -$  & & $107(36)$  &   $0.2(3)$  & $0.12$  \\
$\Delta_n$ & Even Z   & $82, -$  & $50, -$  & & $123(21)$  &   $0.3(1)$  & $0.20$  \\
$\Delta_n$ & Odd Z    & $82, -$  & $50, -$  & & $54(23)$   &   $0.5(1)$  & $0.20$  \\
$\Delta_p$ & Even N   & All      & All      & & $60(4)$    &   $0.8(0)$  & $0.20$  \\
$\Delta_p$ & Odd N    & All      & All      & & $34(4)$    &   $0.7(0)$  & $0.19$  \\
$\Delta_p$ & Even N   & $50, 82$ & $50, -$  & & $-13(51)$  &   $1.5(4)$  & $0.15$  \\
$\Delta_p$ & Odd N    & $50, 82$ & $50, -$  & & $-76(46)$  &   $1.6(4)$  & $0.14$  \\
$\Delta_p$ & Even N   & $82, -$  & $50, -$  & & $143(18)$  &   $0.3(1)$  & $0.17$  \\
$\Delta_p$ & Odd N    & $82, -$  & $50, -$  & & $86(18)$   &   $0.4(1)$  & $0.16$  \\
\colrule
\end{tabular}
\caption{The results for the two-term model from Eq.~(\ref{Eq. ber}), where the RMSE in each region is calculated based on the best local fit. \label{Tab. Ber}}
\end{table}

\begin{table}
\begin{tabular}{l l c c c r  r r  r} 
\colrule
Type               & Nuclei & \multicolumn{2}{c}{Region limits} & & \multicolumn{1}{c}{$c_2$} & \multicolumn{1}{c}{$c_1$} & RMSE     \\                                                   
\cline{3-4} \cline{6-8}
    &    & $N$ & $Z$ && \multicolumn{3}{c}{$\left[ \si{\mega\electronvolt} \right]$}       \\
\colrule
$\Delta_n$ & Even Z   & $50, 82$ & $50, -$  & & $58(5)$    &   $0.7(0)$  & $0.22$  \\
$\Delta_n$ & Odd Z    & $50, 82$ & $50, -$  & & $32(4)$    &   $0.6(0)$  & $0.21$  \\
$\Delta_n$ & Even Z   & $82, -$  & $50, -$  & & $58(5)$    &   $0.7(0)$  & $0.21$  \\
$\Delta_n$ & Odd Z    & $82, -$  & $50, -$  & & $32(4)$    &   $0.6(0)$  & $0.20$  \\
$\Delta_p$ & Even N   & $50, 82$ & $50, -$  & & $61(4)$    &   $0.8(0)$  & $0.19$  \\
$\Delta_p$ & Odd N    & $50, 82$ & $50, -$  & & $35(4)$    &   $0.7(0)$  & $0.18$  \\
$\Delta_p$ & Even N   & $82, -$  & $50, -$  & & $61(4)$    &   $0.8(0)$  & $0.19$  \\
$\Delta_p$ & Odd N    & $82, -$  & $50, -$  & & $35(4)$    &   $0.7(0)$  & $0.17$  \\
\colrule
\end{tabular}
\caption{The same as in Table \ref{Tab. Ber}, but the RMSE in each region is calculated based on the best global fit. \label{Tab. Ber2}}
\end{table}

\begin{figure}
\centering
\includegraphics[width=1\columnwidth]{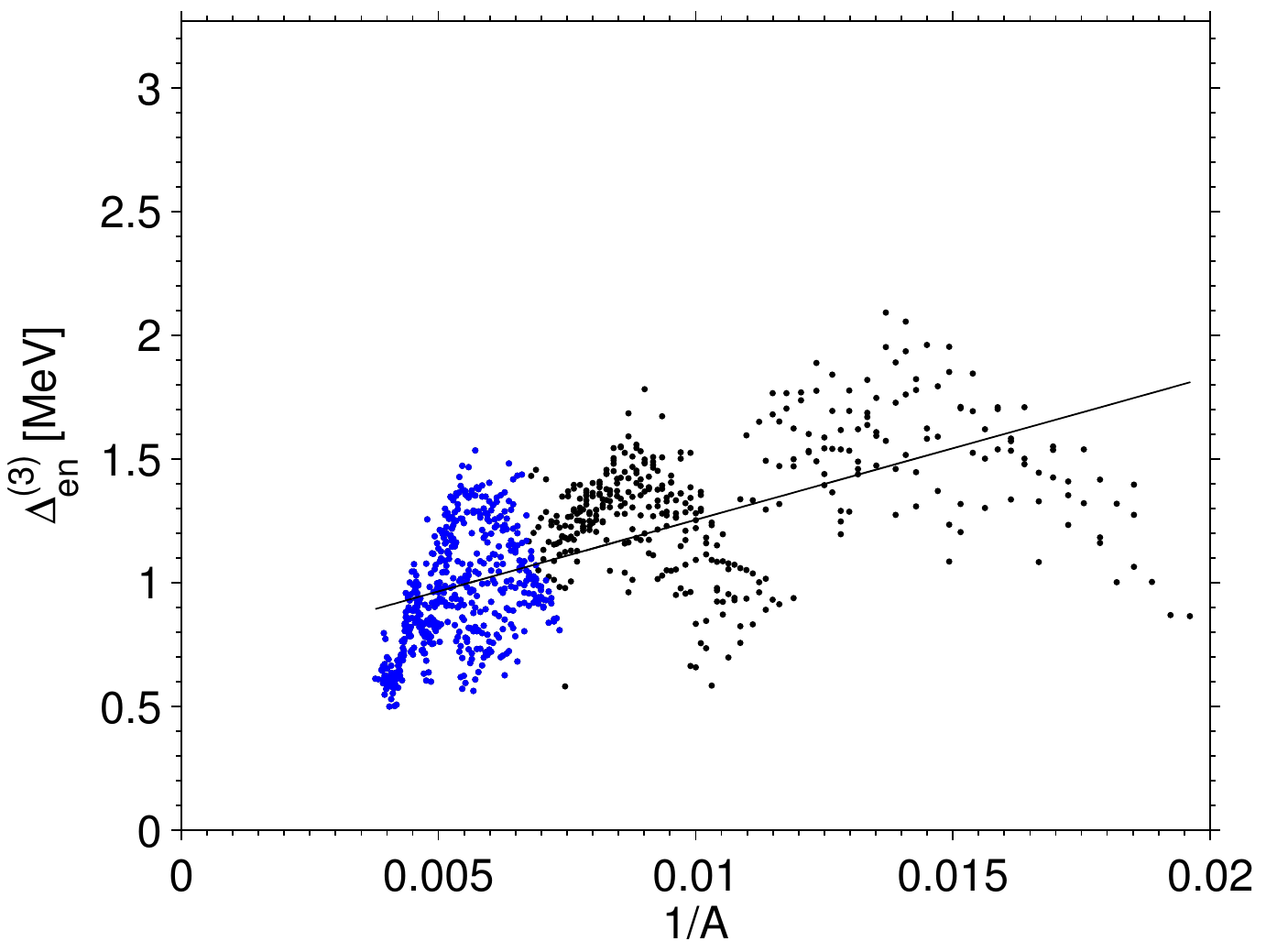}
\caption{(Color online) Best global fit with Eq.~(\ref{Eq. ber}) for even $Z$ nuclei and $A>50$. The nuclei marked in blue are restricted by $82<N$ and $50<Z$. Nuclei influenced by shells or the Wigner effect are not shown. \label{Fig. Ber}}
\end{figure}

This is illustrated in Fig.~\ref{Fig. Ber}, which also shows the
neutron staggering for even nuclei as a function of $A^{-1}$. The line
is the best global fit based on Eq.~(\ref{Eq. ber}), and the nuclei
indicated in blue belongs to the region with $82<N$ and $50<Z$. The
RMSE in Table \ref{Tab. Ber2} is calculated based on a group of
nuclei, such as those marked in blue, with respect to the global fit
indicated by the line. As such it will always be larger than the RMSE
from Table \ref{Tab. Ber}.  From Fig.~\ref{Fig. Ber} it is clear that
Eq.~(\ref{Eq. ber}) does not reproduce the general tendencies observed
in the odd-even staggering. It can, however, still be used as a very
accurate approximation.

Let us also compare briefly to the final expression obtained by Jensen
\textit{et al.} \cite{Aksel} that, in addition to the smooth terms, is
given by
\begin{align}
\tilde{\Delta}(N,Z) 
=& A^{-1/3} \left( 3.68 \pi_n \left(1 - 8.15 \left(\frac{N-Z}{A} \right)^2 \right) \right. \notag \\
&+ \left.\vphantom{\left(\frac{N-Z}{A} \right)^2} 3.78 \pi_p \left(1 - 6.07 \left(\frac{N-Z}{A} \right)^2 \right) \right) \notag \\
&- 43 \frac{|N-Z|}{A} + \frac{\pi_{np} ( 34 - 24 \delta_{N,Z})}{A}, \label{Res. Aksel}
\end{align}
where $\delta_{N,Z}$ is a Kronecker delta. All constants are in units
of MeV.  The last two terms account for Wigner effects, and are not
relevant in the comparison since we do not include nuclei with $N=Z$.
The first terms have the same dependencies as our result in
Eq.~(\ref{Finaleq_heavy}) and can be compared directly, the main
difference being that we have used a three-point mass formula whereas
\cite{Aksel} used four-point mass formulae and evaluated neutron
pairing, proton pairing and neutron-proton pairing separately. The
$\pi_n$ and $\pi_p$ terms are very similar, and the coefficients on
the $(N-Z)^2$ terms are $5-10$\% larger than in the new fits.  This is
presumably due to our substantially larger data set where most added
nuclei have relatively large neutron excess.  The neutron-proton
pairing term has a similar magnitude for mid to heavy mass nuclei in
spite of the different assumed mass dependence.

\begin{figure}
\centering
\includegraphics[width=1\columnwidth]{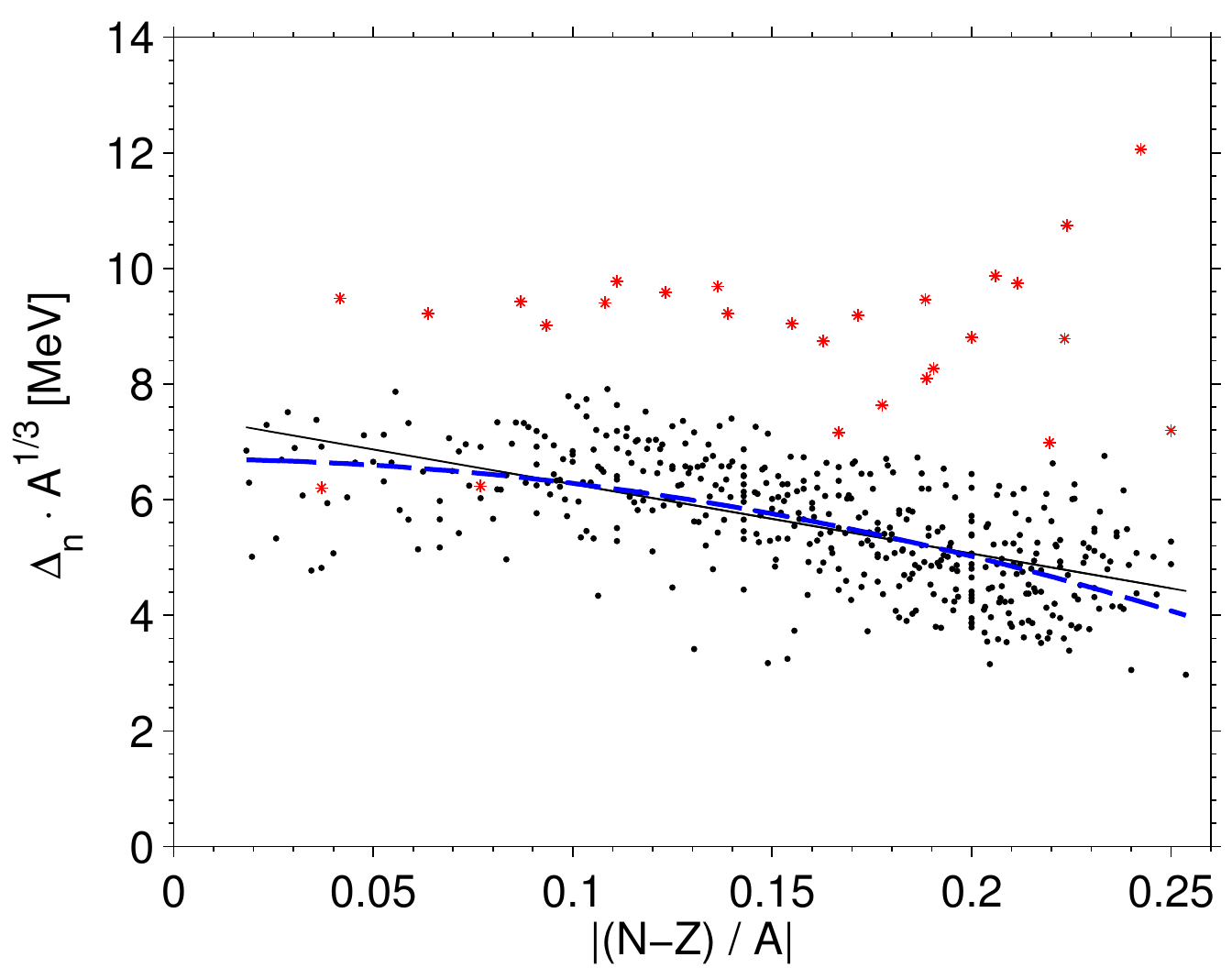}
\caption{(Color online) The equvalent to Fig.~\ref{Iso e-e tot} only with respect to $\left\lvert \frac{N-Z}{A} \right\rvert$. Here the dashed, blue line is the best fit quadratic in neutron excess. \label{Fig. Lin Iso}}
\end{figure}

Finally, the question of whether to use a model which is linear or
quadratic in neutron excess deserves some attention. In Fig.~\ref{Iso
e-e tot} the results were shown in relation to a quadratic neutron
excess for all even-even nuclei with $A>50$, and the best linear fit
was shown with a dashed, blue line. The same figure, but in relation
to a linear neutron excess can be seen in Fig.~\ref{Fig. Lin Iso}. The
result of the best linear fit is $A^{1/3} \Delta_n = -12 \left\lvert
\frac{N-Z}{A} \right\rvert + 7.5 $ with $RMSE = 0.83$, and
$RMSE(\Delta_n) = 0.17$ all in units of $\si{\mega\electronvolt}$. The
dashed, blue line in Fig.~\ref{Fig. Lin Iso} is the best quadratic
fit, the values of which can be found in Table \ref{Tab. Iso}. 
 
The difference between the two descriptions is perhaps 
surprisingly small. Possible reasons could be that both forms
only are approximations to a better generic dependence, or the range
of nuclei is too small to distinguish, or the individual scatter of
points arise from a chaotic behavior prohibiting substantial
improvements in simple fits \cite{mol06}. A significant increase in the number of
measurements of far off stability nuclei would be very helpful to
address these questions.

\section{Conclusion \label{Conclusion}} 

Our phenomenological study of odd-even staggering terms in the nuclear
binding energy makes use of the three-point mass relations, $\Delta_n$
and $\Delta_p$. This second order difference eliminates most smooth
aspects of the binding energy and the liquid drop model was used to
eliminate the remains of the smooth variations. By also avoiding
isotopes with magic numbers, on the $N=Z$ line, or generally very
light only non-smooth, odd-even contributions remains.

The starting point of our description is the trends seen in
Fig.~\ref{3D}. The odd-even configuration of both neutrons and protons
was seen to have an influence on the general scale of the
staggering. The region in question also influenced the result. To
examine, and possibly account, for these observations the nuclei were
separated according to odd-even configuration, and into regions
defined by nuclear shells. 
We find that the difference in $\Delta_n$ for odd and even $Z$ and the
corresponding differences in $\Delta_p$ is naturally described in
terms of a common neutron-proton pairing term. This is in line with
the findings \cite{Aksel,Friedman} made using second order mass
differences. By construction we end up with the same overall mass
dependence for the neutron-proton pairing term as for the other terms,
namely $A^{-1/3}$, where the other works employ $A^{-1}$ and
$A^{-2/3}$. There is as yet no basic theoretical framework that can
explain the neutron-proton pairing systematics \cite{Friedman} so a
closer look at the data may be warranted.

It is well known that the data on odd-even staggering fluctuate
systematically around a power-law fit, and odd-even mass differences
calculated from self-consistent mean field theory \cite{ber09a}
displays a similar behaviour. These systematic deviations are also
clearly visible in Fig.~\ref{3D} and suggest a description in terms of
a dependence on $N-Z$, as attempted earlier \cite{Vog84,Aksel}. It turns
out that the $N-Z$ dependent terms vary in importance as we go from
light to heavy nuclei. For $Z<50$ the results were almost constant,
when considering scattering, but for $50<Z$ there was a clear decrease
as a function of neutron excess. This transition was seen for both
neutrons and protons.

To make the model globally applicable the separated results were
recombined. Since the odd-even $Z$ configuration had a very modest
influence on $\Delta_p$ it was neglected. The neutron-proton pairing
term accounted for the displacement of the odd-$N$ nuclei relative to
the even-$N$ nuclei. Similarly, odd-$Z$ nuclei were displaced for
$\Delta_n$. This resulted in two global expressions for the odd-even
staggering effect as given in Eqs~(\ref{Finaleq_light}) and
(\ref{Finaleq_heavy}).  These odd-even terms have to be included in
phenomenological expressions of the nuclear binding energy where the
largest contributions, liquid drop and shell effects, can be
maintained from previous studies.

The separated and the combined expressions were compared to a two-term
model, with a $A^{-1}$ dependency, and our $(N-Z)^2$ dependence showed
greater accuracy both locally and globally.  The overall root mean
square deviations in all our fits are always (significantly) less than
$0.2$~MeV. There is as yet no theoretical explanation for a systematic neutron
excess dependence of the odd-even staggering for heavy nuclei,
i.e.\ when nuclei fill the large shells, but the fact that the data
follow the fit curves suggests that a --- direct or indirect --- dependence on isospin
projection should be considered on top of the previously included
physical explanations for odd-even staggering \cite{boh75,ber09}.
Finally, we note that calculations including three-nucleon forces have
now been used \cite{Hol13} to study the variation of the odd-even
staggering in the heavy Ca isotopes. The rapid progress in nuclear
theory these years may give a new perspective on this old problem.



\end{document}